\documentstyle[12pt]{article}

\def\.{\!\cdot\!}
\def\:{\cdots}
\def\[{\left[}
\def\]{\right]}
\def\({\left(}
\def\){\right)}

\def\bk#1{\langle#1\rangle}

\def\h{{1\over 2}}

\def\l{\ell}
\def\n{\noindent}

\def\r2{\sqrt{2}}
\def\rs{\sqrt{s}}

\def\+{`$+$'}
\def\-{`$-$'}
\def\lra{\leftrightarrow}
\def\A{{\cal A}}

\def\K{{\cal K}}
\def\bc{\begin{center}}
\def\ec{\end{center}}
\def\bd{\begin{document}}
\def\ed{\end{document}}
\def\be{\begin{eqnarray}}
\def\nn{\nonumber}
\def\ee{\end{eqnarray}}
\def\bn{\begin{enumerate}}
\def\en{\end{enumerate}}
\def\bi{\begin{itemize}}
\def\ei{\end{itemize}}

\begin{document}
\input epsf.tex
\vspace{.5cm}
\rightline{McGill/96-27}
\vspace{2 cm}
\begin{center}
{{\Large\bf Nonabelian Cut Diagrams}\\
\vspace{.5 cm}
C. S. Lam$^{\ \dagger\ *}$\\
\bigskip
{\it Department of Physics, McGill University,\\
3600 University St., Montreal, P.Q., Canada H3A 2T8}}
\end{center}

\vspace{3 cm}

\begin{abstract}
Symmetrization of bosonic wave functions produces bunching and low temperature
phenomena like Bose-Einstein condensation, superfludity, superconductivity,
as well as laser. Since probability is conserved, one might expect such
bunchings at one place would lead to a depletion or a destructive interference
at another place. This indeed takes place in high-energy scatterings, leading
to cancellations between Feynman diagrams with permuted (real or virtual) boson
lines. It is such cancellation that allow the Froissart bound to be obeyed,
and the meson-baryon amplitudes in large-$N_c$ QCD to be consistent.
These lectures review the derivation and some of the applications of
{\it nonabelian cut diagrams}, which are resummation of Feynman diagrams, but
with these interference effects explicitly incorporated. Unlike the traditional
method where cancellations occur between Feynman diagrams, making it
necessary to compute each diagram to subleading orders, destructive
interference has been taken care of from the start in nonabelian cut
diagrams. No further cancellation takes place, so each nonabelian cut diagram
may be computed only in the leading order.
\end{abstract}
\vfill
\n $\dagger$\quad Lectures given at the First Asia Pacific Workshop on
Strong Interactions, Taipei. August 1st to 27th, 1996.

\bigskip
\n $*$\quad Email: Lam@physics.mcgill.ca
\eject

\centerline{\bf TABLE OF CONTENTS}
\bigskip
\begin{enumerate}
\item Preliminaries
\smallskip
\begin{enumerate}
\item[1.] Introduction
\item[2.] Conventions
\item[3.] Feynman rules
\item[4.] Color algebra
\end{enumerate}
\medskip
\item High-Energy Elastic Scattering
\smallskip
\begin{enumerate}
\item[1.] Energy dependence and Regge poles
\item[2.] QED to $O(g^6)$
\end{enumerate}
\medskip
\item Abelian Cut Diagrams
\smallskip
\begin{enumerate}
\item[1.] Factorization formula
\item[2.] Sum of QED $s$-channel-ladder diagrams
\item[3.] Other sum rules
\end{enumerate}
\medskip
\item  Nonabelian Cut Diagrams
\smallskip
\begin{enumerate}
\item[1.] Multiple commutator formula
\item[2.] Folding formula
\end{enumerate}
\medskip
\item Quark-Quark Scattering to $O(g^6)$
\smallskip
\begin{enumerate}
\item[1.] Color factors
\item[2.] Sum of Feynman diagrams
\item[3.] Sum of nonabelian cut diagrams
\end{enumerate}
\medskip
\item Multiple Reggeons and QCD
\smallskip
\begin{enumerate}
\item[1.] The reggeized factorization hypothesis
\item[2.] BFKL equation and unitarity
\item[3.] Nonabelian cut diagrams vs Feynman diagrams
\item[4.] $s$-channel-ladder diagrams
\end{enumerate}
\medskip
\item References

\end{enumerate}

\vfill\eject

\begin{center}
\section{PRELIMINARIES}
\end{center}
\bigskip
\subsection{Introduction}
The symmetrization of bosonic wave functions leads to an effective
attraction. At low temperatures this constructive interference gives rise to
the Bose-Einstein
condensation, superfluidity of Helium, and superconductivity, as well as laser.

Probability is conserved. If there is a concentration of wave function at
one place then there is presumably a depletion at another. How can
one observe this depletion, or destructive interference? It turns out that
this can be seen in high energy processes. The effect is not
as dramatic as the low-temperature constructive interference processes, but
its presence is absolutely necessary to render theoretical consistency,
as we shall see.

A general discussion of this interference can be found in Secs.~3 and 4.
The interference effect can be incoporated in our calculations by using the
the {\it multiple commutator formula} \cite{1} and the {\it nonabelian
cut diagrams} \cite{2}. These cut diagrams may be
considered
as resummations of Feynman diagrams where the interference of the identical
bosons in the diagrams, whether real or virtual, has been explicitly
accounted for at the very first step, the step of drawing the diagrams
before any computation begins. With the interference thus included,
subsequant cancellations are not needed and will not occur.
In contrast, if as usual
Feynman diagrams are used, substantial amount of delicate cancellations
between diagrams will occur at a later stage as a result of such destructive
interference. This is often much more difficult to calculate as will be
elaborated below.

A fairly dramatic example of this can be found in the QCD theory with a large
number of color ($N_c\gg 1)$.
Its $n$-meson tree amplitude in the one-baryon sector, {\it viz.,}
meson-baryon scattering producing $n-1$ mesons in the final state,
depends on $N_c$ as $N_c^{1-\h n}$\cite{3}. In particular, taking $n=1$, this
shows that the meson-baryon
Yukawa coupling constant go like $\sqrt{N_c}$. Since the baryon propagators
are of order unity, individual Feynman tree diagrams will go like
$N_c^{\h n}$, which is $n-1$ powers of $N_c$ larger than what the full
amplitude should be.
Unless a huge cancellation occurs when we sum up the $n!$ Feynman diagrams,
the theory will not be consistent \cite{4}! Such cancellation indeed occurs, it
is a manifestation of the destructive interference mentioned earlier, but to
see how that works directly using Feynman diagram is very difficult task.
Not so with the nonabelian cut diagrams \cite{5}
because the interference effects leading to the cancellations
have already been built in. We will not
pursue this subject further here because it is a bit outside of the scope
of this workshop.

What we will discuss is destructive interference in high-energy elastic
scattering taking place in loops,  where identical bosons are present in
the intermediate states for the interference to take place. Individual Feynman
diagrams with many loops often contain
high powers of $\int_{m^2/\sqrt{s}}d\omega/\omega\sim \ln s$ \cite{6,7,8,9}. If
not cancelled,
the Froissart bound would be violated. When Feynman diagrams are summed,
such offending $\ln s$ do get cancelled \cite{6} as a result of the destructive
interference mentioned earlier. However, it is difficult to see how this
happens in traditional calculations
for two reasons. First, one has to do a lot of virtual work
to compute terms that eventually get cancelled. Second and more seriously,
such computations can usually be carried out only in the leading-log
approximation. If these leading powers of $\ln s$ are cancelled out, then one
needs to compute the subleading, the sub-subleading (etc.) terms in order to
obtain
a non-zero finite result at the end. Such tasks are often impossible except
in very low orders. Again, the nonabelian cut diagrams bypass these
difficulties \cite{2,10,11} because the destructive interference effect has
been taken care of before any computation even begins.

\subsection{Conventions}
We shall adopt the following conventions for the metric $g^{\mu\nu}$, the Dirac
matrix $\gamma^\mu$, and the Dirac spinor $u$:
\begin{eqnarray}
g^{\mu\nu}&=&(+---)_{diag}\nonumber\\
\{\gamma^\mu,\gamma^\nu\}&=&2g^{\mu\nu}\nonumber\\
\bar u u&=&2M\ .
\end{eqnarray}
The plane-wave states are normalized covariantly,
\begin{eqnarray}\bk{{\vec p\,}'\,|\vec p\,}=(2\pi)^3(2p^0)\delta^3(\vec p\,'-\vec p)\
,\end{eqnarray}
the $S$-matrix is related to the covariant $T$-matrix by
the formula
\begin{eqnarray}S_{fi}={\bf 1}_{fi}+(2\pi)^4i\delta^4(\cdots)T_{fi}\ ,
\end{eqnarray}
where $\delta^4(\cdots)$ is the overall energy-momentum conservation
$\delta$-function, and the unitarity relation is given by
\begin{eqnarray}-{i\over 2}\(T_{fi}-T^*_{fi}\)=
{\rm Im}\(T_{fi}\)=\h\sum_n\int T^*_{fn}d\Phi_n
T_{ni}\ ,
\end{eqnarray}
in which the expression for the $n$-body invariant phase space is
\begin{eqnarray}\int d\Phi_n=\int\prod_{i=1}^n\({1\over (2\pi)^3}{d^3p_i'\over 2{p_i'}^0 }\)
(2\pi)^4\delta^4(\cdots)\ .
\end{eqnarray}

For elastic scattering with initial momenta $p_i$ and final momenta $p_i'$,
momentum conservation dictates that $p_1+p_2=p_1'+p_2'$. The Mandelstam
variables are $s=(p_1+p_2)^2, t=(p_1-p_1')^2$, and $u=(p_1-p_2')^2$; they
are related by $s+t+u=2(M_1^2+M_2^2)$. $\rs$ is the centre-of-mass energy, and
$\Delta=\sqrt{-t}$ is the momentum transfer.
We shall be concerned with situations where $s$ is much larger than $-t$ and
$M_i^2$. The elastic cross-section is related to the elastic
amplitude $T_{fi}$ by
\begin{eqnarray}\int d\sigma=\int {1\over 2s}|T_{fi}|^2d\Phi_2={1\over 16\pi s^2}\int
|T_{fi}|^2dt\ ,
\end{eqnarray}
and the total cross-section is given by unitarity to be
\begin{eqnarray}\sigma_{tot}={1\over s}{\rm Im}(T_{ii})\ .
\end{eqnarray}

\subsection{Feynman Rules}

The Feynman rules for the $T$-matrix elements that will be needed
are the following.
\medskip
\begin{itemize}
\item Quark propagator: $(M+\gamma\.p)/(p^2-M^2+i\epsilon)$
\item Gluon propagators in the Feynman gauge:
$-g^{\mu\nu}/(p^2+i\epsilon)$
\item Quark-gluon vertex: $g t_a \gamma^\alpha$
\item Triple-gluon vertex: $g(if_{abc})\{g^{\alpha\beta}(p_1-p_2)^\gamma
+g^{\beta\gamma}(p_2-p_3)^\alpha+g^{\gamma\alpha}(p_3-p_1)^\beta\}$
\item Four-gluon vertex: $g^2\{f_{abe}f_{cde}(g^{\alpha\gamma}g^{\beta\delta}
-g^{\alpha\delta}g^{\beta\gamma})
+ f_{ace}f_{bde}(g^{\alpha\beta}g^{\gamma\delta}-
g^{\alpha\delta}g^{\beta\gamma})
+\hfill\break
f_{ade}f_{bce}(g^{\alpha\beta}g^{\delta\gamma}-g^{\alpha\gamma}g^{\delta\beta}
)\}$
\item External wave functions: $\epsilon^*_\lambda(p), \epsilon_\lambda(p),
\bar u_\lambda(p), u_\lambda(p)$
\item Loop integration: $i\int d^4q/(2\pi)^4$
\item Other factors: an overall minus sign, a minus sign per fermion loop, and
symmetry factors.
\end{itemize}

\subsection{Color Algebra}

The $SU(N_c)$ color matrices $t_a$
obeys the commutation relation
\begin{eqnarray}[t_a,t_b]=if_{abc}t_c\ .
\end{eqnarray}
The structure constants satisfy the sum rules
\begin{eqnarray}f_{abc}f_{abd}=N_c\delta_{cd},\quad
i^3f_{adg}f_{bed}f_{cge}=i{N_c\over 2}f_{abc}\equiv icf_{abc}\ .
\end{eqnarray}
See Fig.~1 for a graphical representation of (1.8) and (1.9).

\begin{figure}
\vskip -0 cm
\centerline{\epsfxsize 3 truein \epsfbox {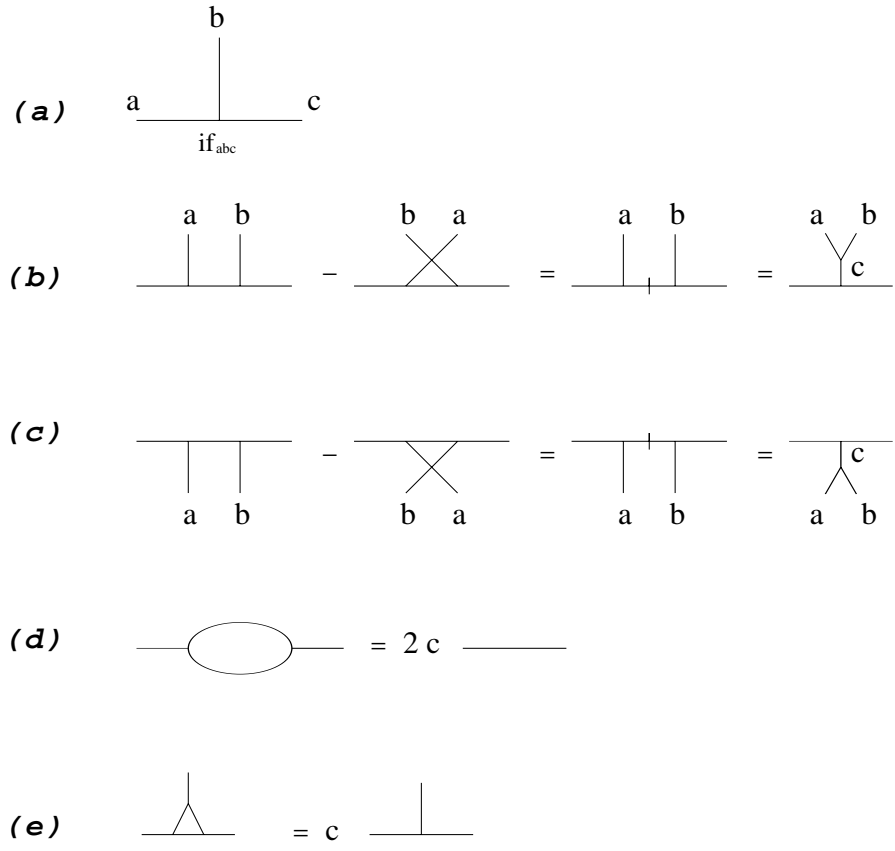}}
\nobreak
\vskip -6 cm\nobreak
\vskip 1.5 cm
\caption{Graphical representation of the color-matrix relations (1.8) and
(1.9).} \end{figure}
\newpage

\begin{center}
\section{HIGH-ENERGY ELASTIC SCATTERING}
\end{center}
\bigskip
\setcounter{equation}{0}
\subsection{$s$-Dependence and Regge Poles}

\begin{figure} [h]
\vskip -9 cm
\centerline{\epsfxsize 3 truein \epsfbox {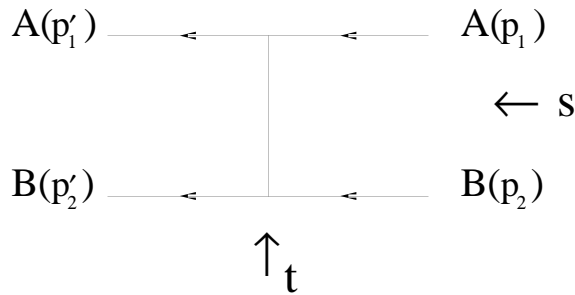}}
\nobreak
\vskip -3 cm\nobreak
\vskip 0.5 cm
\caption{An elastic scattering process $A+B\to A+B$.}
\vskip 1cm
\end{figure}

Consider the tree diagram in Fig.~2 for the elastic process $A+B\to A+B$.
For simplicity we shall assume $A$ and $B$ to have the same mass.
If all the particles are scalars
then the T-matrix amplitude is $-g^2/t$ and is energy independent.
On the other hand,
if the exchange particle $C$ is a photon, then the vertices add on
an additional
factor $\simeq -(2p_1)\.(2p_2)\simeq -2s$, now the amplitude
grows linearly with $s$. In general, the amplitude grows like $s^\l$ if the
spin of the exchanged particle $C$ is $\l$. To see that, analytically
continue the amplitude to the physical region of the $t$-channel process
$B+\overline{B}\to \overline{A}+A$, where $t>0, s<0, u<0$.
In its CM system, the total angular momentum is $\l$, so the amplitude is
proportional to $P_\l(\cos\theta_t)$, where $\theta_t$ is the CM scattering
angle for this process and it is related to $s$ and $u$ by
\begin{eqnarray}s=-{t\over 2}(1-\cos\theta_t)\ ,\quad u=-{t\over 2}(1+\cos\theta_t)\ .
\end{eqnarray}
We can now analytically continue back to the physical region of the $s$-channel
process and examine
the limit when  $s\to\infty$. Since
\begin{eqnarray}P_\l(z)\sim z^\l\ ,\quad ({\rm Re}\l>-\h)
\end{eqnarray}
for large $z_t\equiv -\cos\theta_t\sim s$, the amplitude grows like $s^\l$ as
claimed.

What if we have a complicated diagram so that the $t$-channel
angular momentum is not fixed? Then the amplitude is given by a partial-wave
expansion
\begin{eqnarray}
T(s,t)=\sum_\l a_\l(t)(2\l+1)P_\l(z_t)\ ,\end{eqnarray}
say with some `mean' angular momentum
$\overline{\l}\equiv \alpha(t)$. If the spread of angular momentum is small,
one might expect simply by interpolation that
the amplitude still grows like $s^{\alpha(t)}$.
This naive expectation will be justified below even without this narrow-spread
assumption. The function $\alpha(t)$ is known as a {\it Regge trajectory}.

Let us assume $a_\l(t)$ to be an analytic
function of $\l$, dying off sufficiently fast at infinity to enable
the Sommerfeld-Watson representation to be used. We can then replace the sum
(2.3) by a contour integral wrapping around the positive real $\l$-axis,
\begin{eqnarray}T(s,t)=-{1\over 2i}\int_\Gamma
d\l(2\l+1)a_\l(t){P_\l(-z_t)\over \sin\pi \l}\ ,
\end{eqnarray}
as shown in Fig.~3.

\begin{figure}
\vskip -0 cm
\centerline{\epsfxsize 3 truein \epsfbox {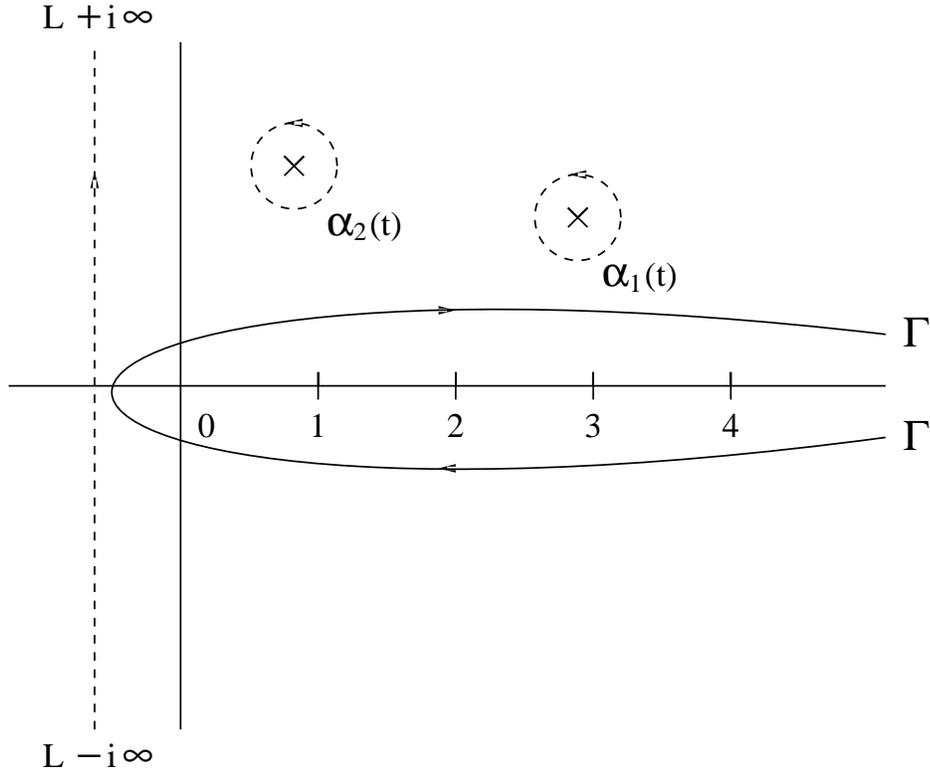}}
\nobreak
\vskip -6 cm\nobreak
\vskip 1.5 cm
\caption{Various contours used for the Sommerfeld-Watson representation.}
\end{figure}

Suppose $a_\l(t)$ has (Regge) poles located at
$\l=\alpha_i(t)$ with residue $\beta_i(t)/(2\alpha_i(t)+1)$.
Opening up the contour $\Gamma$ and moving it to the left of the poles,
we get
\begin{eqnarray}T(s,t)=-{1\over 2i}\int_{L-i\infty}^{L+i\infty}
d\l(2\l+1)a_\l(t){P_\l(-z_t)\over \sin\pi \l}
-\sum_i{\pi\beta_i(t)\over
\sin\pi\alpha_i(t)}P_{\alpha_i(t)}(-z_t) \ ,
\end{eqnarray}
where $L$ is smaller than the real part of all $\alpha_i(t)$.
The high-energy behavior is now dominated by the rightmost Regge pole
$\alpha(t)$ to be
\begin{eqnarray}(-z_t)^{\alpha(t)}\simeq (2s/t)^{\alpha(t)}\ ,
\end{eqnarray}
agreeing with the naive expectation.

In trying to get to this quickly I have glossed over a number of fine points.
I have not discussed why $T(s,t)$ is analytic but that can be established at
least in perturbation theory. I have ignored the spins of the colliding
particles which complicate matter but pose no essential
difficulty.  Helicity is conserved
at high energies and the dominant amplitude is helicity independent,
so for elastic scattering everything is essentially the same as if
the beam and target particles were without spin anyway.
Another caveat is that the asymptotic behavior (2.2) is valid only
when Re$(\l)>-\h$ so one might worry about the general validity of (2.6).
It turns out that this can always be fixed up
to render (2.6) valid.

I have also not discussed why it is reasonable to assume $a_\l(t)$
to be analytic in $\l$. In fact, this is not even true unless
the {\it signature} factor is taken into account.

To see what signature is let us suppose the particle $A$ is its own
antiparticle:
$\overline{A}=A$. Then there is a forward-backward symmetry for the $t$ channel
process $\overline{B}+B\to \overline{A}+A$ so that the amplitude $T(s,t)$
is $s\lra u$ symmetric. From (2.1) and (2.3), we see that only even angular
momenta $\l$ are present; $a_\l(t)$ would be zero for odd $\l$.
One would therefore expect that only even $\l$ values are smoothly connected
by the function $a_\l(t)$.

This example suggests that we should always decompose an amplitude
$T(s,t)$ into its $s\lra u$ symmetric/antisymmetric parts $T_\pm(s,t)$.
Both $a_{\pm\l}(t)$ are analytic but they may have different singularities, so
the Regge trajectories for these two will generally be different.
These two kinds of Regges trajectories are said to have
even/odd signatures. When we replace $T$ in (2.3) by $T_\pm$,
the right-hand side should be replaced by $P_\l(z_t)\pm P_\l(-z_t)$,
so the last factor in (2.5) and (2.6) become
\begin{eqnarray}{1\over \sin\pi\alpha_i(t)}
[P_{\alpha_i(t)}(-z_t)\pm P_{\alpha_i(t)}(z_t)]&\to& {1\over\sin\pi\alpha_i(t)}
[(-z_t)^{\alpha_i(t)}
\pm (z_t)^{\alpha_i(t)}]\nonumber\\
&\to& \({s\over t}\)^{\alpha_i(t)}{1\over\sin\pi\alpha_i(t)}
[1\pm e^{\pi i\alpha_i(t)}]\ .
\end{eqnarray}
The factor $(1\pm e^{\pi i\alpha(t)})/\sin\pi\alpha(t)$
blows up at even/odd integer angular momenta $J$.
This pole at $t=m_J^2$ corresponds to a resonance in the $t$-channel
with mass $m_J$ and spin $J$, so
the Regge trajectory $\alpha(t)$ may be regarded as a trajectory
connecting particles in the $J$ -- $m_J$ plane.
Extrapolating this trajectory to negative $t$, it will determine
the high-energy behavior
of the $s$-channel amplitude, and this is the power of the Regge theory!
However, this would work only if the singularities of $\alpha_\l(t)$
in $\l$ are indeed poles, an assumption that requires explicit verification
and is not automatically guaranteed. In fact, as we shall see,
the pole assumption appears to be true for the gluon trajectory but not
the Pomeranchuk trajectory that is discussed see below.
B
Experimentally the total cross-section grows with energy, thus from (1.7) and
(2.7) it follows that the leading regge intercept is at least 1, $\alpha(0)\ge
1$. However, the
Froissart bound forbids the total cross-section to grow faster than
$(\ln s)^2$ asymptotically,
so this leading singularity, called the {\it Pomeranchuk} trajectory or the
{\it Pomeron} for short, must have an unit intercept $\alpha(0)=1$, but then it
cannot be a simple pole for otherwise the total cross-section would be a
constant. It may be a double pole
or something more complicated. The usual view is that in QCD it is a composite
object made up of two and more reggeized gluons.

\subsection{QED to $O(g^6)$}

Let us now turn to actual calculations of  high energy
processes in QED. We will follow the book of Cheng and Wu [6]
from which one can find more details, as well as references to the original
literature.
We will concentrate specifically on electron-electron elastic scattering in the
following.

The incoming beams will be assumed to move in the $z$-direction.
At high energy it is convenient to use the lightcone coordinates
$a_\pm=a^0\pm a^3$, in terms of which the dot product
of two four-vectors $a=(a_+,a_-,a_\perp)$ and $b=(b_+,b_-,b_\perp)$ is given by
$a\.b=\h(a_+b_-+a_-b_+)-a_\perp\.b_\perp$. The momenta of the incoming
beams in the CM system are $p_1=(\rs,0,0)$ and $p_2=(0,\rs,0)$, with masses
ignored. We will make it a practice to draw $p_1$
on top and $p_2$ at the bottom in all the scattering diagrams.

We shall label
the loop momenta by $q_i$ and define the scaled \- momenta to be $x_i=q_{i-}/\rs$. The measure for loop integration in lightcone variables is
\begin{eqnarray}{i\over (2\pi)^4}d^4q&=&{i\over 8\pi^2}dq_+dq_-{1\over (2\pi)^2}
d^2q_\perp
={1\over 4\pi}\[{\rs\over -2\pi i} dq_+\]dx\[{1\over (2\pi)^2}
d^2q_\perp\]\nonumber\\
&\equiv& {1\over 4\pi}[Dq_+]dx[Dq_\perp] \ .
\end{eqnarray}

The momenta along the upper electron line are   $p_1+Q$, with
$Q$ being some combination of $q_i$ (see Figs.~4 and 5 for illustrations).
In the leading-log approximation when $|\Delta|=\sqrt{-t}$ is fixed and
$s\to\infty$, we can approximate the electron propagator by
\begin{eqnarray}{M+\gamma(p_1+Q)\over (p_1+Q)^2-M^2+i\epsilon}\simeq{M+\gamma p_1
\over 2p_1\.Q+i\epsilon}\ .
\end{eqnarray}
Since
\begin{eqnarray}M+\gamma p_1=\sum_\lambda u_\lambda(p_1)\overline
u_\lambda(p_1)\
, \quad \overline u_{\lambda'}(p_1)\gamma^\mu u_\lambda(p_1)=(2p_1^\mu)
\delta_{\lambda'\lambda}\ ,\end{eqnarray}
we may simply drop all the $\gamma$ matrices and spinor numerators, and
consider the
vertices to be $2p_1^\mu$.   The same is true for the lower electron line where
vertices can be taken to be $2p_2^\nu$. Physically, this expresses the simple
fact that the large electric current carried by an electron at high energy
is a result of its fast motion, with its magnetic moment  contributing
only a negligible amount.
For the same reason the current of a spin-1 particle can be taken to be
$2p^\mu$, which is valid even when that is the color current imbedded
in a triple-gluon vertex in QCD, because the other two terms
in the vertex are of order
$\Delta/\rs$ and can be neglected.

\bigskip

The procedure to follow to compute the high-energy amplitude is as follows:
\smallskip
\begin{enumerate}
\item Use residue calculus and flow diagrams to compute the \+ integration.
\item Then obtain powers of $(\ln s)$ from the \- integrations.
\item The $q_{i\perp}$ integrations are never explicitly carried out.
\end{enumerate}

\subsubsection{The \+ integration}

Let us consider step 1 in more detail.

First of all,  $a\.b$ is linear in $a_+$ and $b_+$, hence the
denominators of all propagators are linear in $q_{i+}$. This then provides
one simple pole per propagator for  the \+ integrations.
The integration contour is originally along the real $q_{i+}$ axis, but we will
close it  with an extra half circle on the upper plane or the
lower plane. By using $-q_{i+}$ as the integration variable if necessary,
we can and will always choose to close the contour in the lower plane.
The result of the integral is therefore equal to $-2\pi i$
times the residue at the pole, summed over all poles in the lower plane.
The important question to decide then is which are the poles
in the lower plane, and which are the poles in the upper plane.
This clearly depends on the sign of the \- components, but they  are
integration variables and hence vary. The
 {\it flow diagrams} are invented to get the various possibilities
sorted out [6].

\begin{figure}[h]
\vskip -0 cm
\centerline{\epsfxsize 3 truein \epsfbox {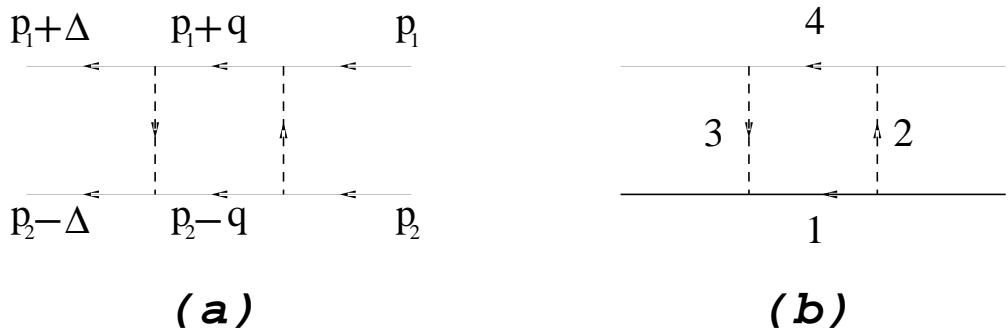}}
\nobreak
\vskip -10 cm\nobreak
\vskip 1.5 cm
\caption{A one-loop Feynman diagram (a) and its flow diagram (b).}
\end{figure}

\begin{figure}
\vskip -0 cm
\centerline{\epsfxsize 3 truein \epsfbox {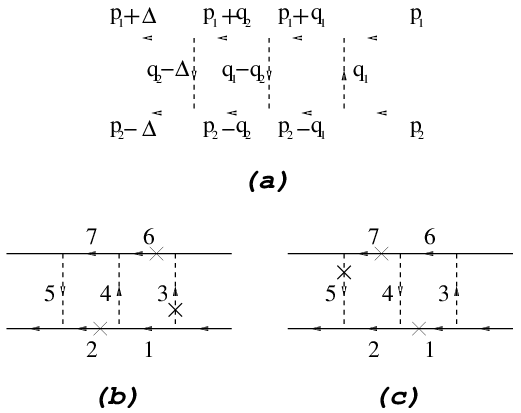}}
\nobreak
\vskip -8 cm\nobreak
\vskip 1.5 cm
\caption{A two-loop Feynman diagram (a) and its flow diagrams (b) and (c).}
\end{figure}

Figs.~4(a) and 5(a) are Feynman diagrams and 4(b), 5(b), and 5(c) are
flow diagrams.
The latter are diagrams indicating flows of the \-
components of the momenta, subject to momentum conservation at each vertex.
Arrows going one way (clockwise or counter clockwise)
correspond to  positive \- components, and arrows going the other way
correspond to negative \- components. Which is which does not matter
because we can always change the integration variable from $q_{i+}$
to $-q_{i+}$ to compensate our choice.

Note that we have not included in Figs.~4 and
5 any flow diagram in which the arrows go  around in the same direction in a
close loop,
for this gives rise to poles all in the same half plane so the \+ integration
would vanish. In making up flow diagrams, such
closed loops are always excluded. In particular, this means that
at the injection vertex where $p_2$ enters, and at the ejection
vertex where $p_2'$ leaves, the  arrows are always arranged
in the way shown in these diagrams.  This rule determines uniquely the flow
pattern of a one-loop
diagram, as shown in 4(b), but it does not do so
for diagrams of two or more loops, as the
flow direction at the boundary of two loops can always be reversed.
For two loops we have two flow diagrams, 5(b) and 5(c),
and for more loops there are more flow diagrams.

Since we would like to have as few poles as possible in the lower
plane, we would
choose those directions in every loop with less arrows to correspond to
poles in the lower plane.
These  are indicated by a cross (x) in the flow
diagrams.

In the case of the right-hand loop in 5(b), there are two arrows along
each direction, so it would seem to have two equally valid choices of poles
for the lower plane. However, the ones indicated are preferred for two reasons.

First and foremost, it is never a good idea to choose a pole at the boundary
of two loops. A pole within loop $i$ determines the value of $q_{i+}$ by
the location of the pole, but a pole at the boundary of loops $i$ and
$j$ determines
only the combination $q_{i+}+q_{j+}$ which is less convenient.

There is a second reason for that choice. With the approximation
(2.9), $q_{i+}$ never appears on a propagator along the top line,
so apparent poles there are actually absent.
With that, the choice indicated actually has only one
pole per loop.

\subsubsection{The \- integration}

The inverse propagators along the upper electron line is given by (2.9)
to be $sX$, where $X=Q_-/\rs$ is some linear combination of $x_i$, and
each $x_i$ in a flow diagram is by definition positive.
The $\ln s$ factors are typically obtained from $\int_\epsilon dx_i/x_i$,
where $\epsilon\sim \Delta^2/s$ represents a cutoff of $x_i$ above
which the approximation (2.9) is valid. For that reason, we can
usually ignore all $x_i$ compared to 1 in the leading-log approximation.

To proceed further it is convenient to adopt some shorthand notations.
We will write $a_m=k_{m\perp}^2$ if $k_m$ is the
four-momentum of the $m$th line. If its pole
is in the upper plane, we let $d_m^{-1}$ to be its propagator. If it is
in the lower plane, we let
$d_m^{-1}$ to be its residue multiplied by $\rs$. It is understood that
all of them are evaluated at the lower-plane pole positions,
and the minus sign in the gluon propagator
is included in this factor.

From (2.8), we see that
the T-matrix after the $Dq_{i+}$ integrations is given by the formula
\begin{eqnarray}T=-\int\[\prod_{i=1}^\ell {dx_i\over 4\pi}Dq_{i\perp}\]{N\over D}\ .
\end{eqnarray}
Here $\ell$ is the number of loops in the diagram, $D=\prod_md_m$,
and $N$ is the numerator factor coming from products of vertex factors.
The lower limits of $x_i$ can be taken to be $\epsilon=\Delta^2/s$.
The minus sign comes from the overall sign factor in the last item of
Sec.~1.3.

\subsubsection{Fig.~4}

From Figs.~4(a) and 4(b), we see that
the pole is located at $q_+=a_1/(1-x)\rs\simeq a_1/\rs$, with
$d_1=(1-x)\simeq 1$. The rest of the $d$-factors are $d_2=a_2, d_3=a_3,
d_4=sx$, so $D\simeq 1\.a_2\.a_3\.sx$. The numerator is $N=g^4[(2p_1)\.(2p_2)]^2
=g^4(4s^2)$, so
the T-matrix of this diagram is obtained from (2.11) to be
\begin{eqnarray}T&=&- {g^4s\over\pi}\int
Dq_\perp{1\over a_2a_3}\int_{\Delta^2/s} {dx\over
x}\nonumber\\
&=&-{s\over\pi}(\ln s)g^2 I_2(\Delta)\ ,
\end{eqnarray}
where

\begin{eqnarray}I_n(\Delta)=
\int\[\prod_{i=1}^n{d^2q_{i\perp}\over (2\pi)^2}\]
(2\pi)^2\delta(\sum_{i=1}^nq_{i\perp}-\Delta)\ .
\end{eqnarray}

In terms of convolution defined by
\begin{eqnarray}
(F*G)(\Delta)\equiv \int {d^2q_\perp\over(2\pi)^2}F(\Delta-q_\perp)G(q_\perp)\
,\end{eqnarray}
the function $I_n$ is the $n$th-power convolution of $I_1$
with itself, which we write as
\begin{eqnarray}
I_n(\Delta)=(*I_1)^n(\Delta)\ .\end{eqnarray}
The impact-parameter-space function $\tilde F(b)$
is the Fourier transform of $F(q_\perp)$,
\begin{eqnarray}
\tilde F(b)\equiv\int{d^2q_\perp\over(2\pi)^2}F(q_\perp)\exp(-iq_\perp\.b)
\ .\end{eqnarray}
Such functions are convenient because in this space
convolutions turn into ordinary products in
the usual way: $[\widetilde{F*G}](b)=\tilde F(b)\tilde G(b)$.
Hence
$\tilde I_n(b)=[I_1(b)]^n.$

\subsubsection{Fig.~5}

First consider flow diagram 5(b). The poles are located at
$q_{2+}=a_2/(1-x_2)\rs\simeq a_2/\rs$, with $d_2\simeq 1$,
and $q_{1+}=a_3/x_1\rs$ with $d_3=x_1$. The rest of the $d$-factors
are $d_1=(p_2-q_{1})_+(p_2-q_{1})_--a_1\simeq -a_3/x_1-a_1\simeq -a_3/x_1,
  d_4=-(q_1-q_2)_+(
q_1-q_2)_-+a_4\simeq -(x_1-x_2)a_3/x_1+a_4, d_5=a_5, d_6=sx_2, d_7=sx_1$.
Hence $D=1\.(-a_3/x_1)\.a_3\.\{[x_2a_3+x_1(a_4-a_3)]/x_1\}\.x_1\.sx_2\.sx_1
\simeq -s^2x_2a_3a_5[x_2a_3+x_1(a_4-a_3)]$.
The numerator is $N=g^6(2s)^3$, so the contribution from 5(b) is

\begin{eqnarray}T_b&=&{g^6s\over 2\pi^2}\int Dq_{1\perp}Dq_{2\perp}{1\over a_3a_5}
\int {dx_1dx_2\over x_2[x_2a_3+x_1(a_4-a_3)]}\nonumber\\
&=&{g^6s\ln s\over 2\pi^2}\int Dq_{1\perp}Dq_{2\perp}{\ln(a_4/a_3)\over a_3a_5(a_4-a_3)}\ .
\end{eqnarray}

The flow diagram 5(c) gives identical result so the total is
\begin{eqnarray}T&=&T_b+T_c=-{s\ln s\over \pi^2}g^6J_3(\Delta)\ ,\nonumber\\
J_3(\Delta)&\equiv&
\int Dq_{1\perp}Dq_{2\perp}{\ln(a_4/a_3)\over a_3a_5(a_4-a_3)}
\ .
\end{eqnarray}

\subsubsection{2nd, 4th, and 6th order QED diagrams}

QCD diagrams up to $O(g^6)$ are shown in Fig.~6, but for now we
are interested only in those that contribute to QED electron scatterings.
These are diagrams A, B1, B2, C15--C20.

\begin{figure}
\vskip -0 cm
\centerline{\epsfxsize 3 truein \epsfbox {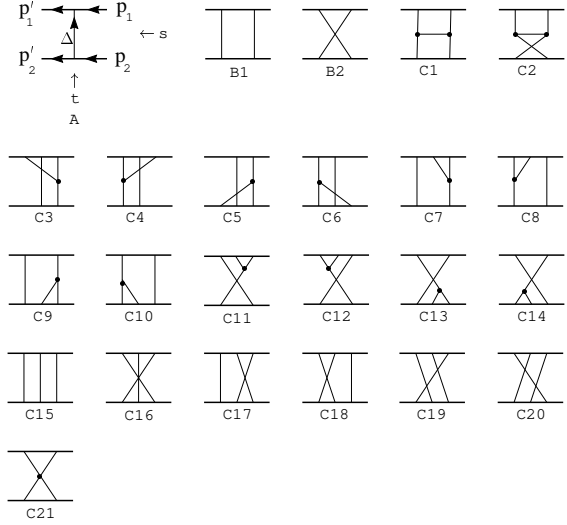}}
\nobreak
\vskip -2 cm\nobreak
\vskip -.5 cm
\caption{QCD scattering diagrams to $O(g^6)$. The QED diagrams come from
the following subset: A, B1, B2, and C15 to C20.}
\end{figure}

B1 and C15 were computed above,
others can all be computed in a similar manner. The contribution to
$T/(2s)$ from each diagram is \cite{6}:

\def\ls{\ln s}
\def\lsb{(\ln s)}
\def\lsi{\ln\(se^{-\pi i}\)}
\def\lssi{\ln^2\(se^{-\pi i}\)}
\def\lss{\ln^2s}
\def\lssb{(\ln^2\!s)}
\begin{eqnarray}
A&=&-g^2I_1\nonumber\\
B1&=&- {\lsi\over 2\pi}g^4 I_2\nonumber\\
B2&=&+{\lsb\over 2\pi}g^4 I_2\nonumber\\
C15&=&-{\lsb\over 2\pi^2}g^6J_3\nonumber\\
C16&=&-{\lsb\over 2\pi^2}g^6J_3 \nonumber\\
C17&=&+{\lsb\over 4\pi^2}g^6(J_3+\pi iI_3)=C18\nonumber\\
C19&=&+{\lsb\over 4\pi^2}g^6(J_3-\pi iI_3)=C20\ .
\end{eqnarray}

The total contribution up to $O(g^6)$ is obtained by adding up all the
expressions in (2.19). The interesting, but sad thing, is that the
$(\ln s)$ contributions all add up to cancel one another. It turns out
that the $O(1)$ contribution from $B1+B2$ is shown correctly in (2.19),
but that of the sum of C15 to C20 is not. So in order to get a nonzero
sum for those a much more difficult calculation accurate to $O(1)$
would have to be carried out.

\begin{center}
\section{ABELIAN CUT DIAGRAMS}
\end{center}
\setcounter{equation}{0}
\bigskip
\subsection{Factorization Formula}

We want to deal with tree diagrams like Fig.~7,
in which the momentum $p$ is much larger than all the momenta $q_i$,
so that the following approximation for the denominator
of propagators is valid:
\begin{eqnarray}(p+\sum_{j=1}^iq_j)^2-m^2\simeq 2p\.\sum_{j=1}^iq_j
\equiv \sum_{j=1}^i\omega_j
\end{eqnarray}
 This is the same approximation used earlier in (2.9).

\begin{figure}
\vskip -0 cm
\centerline{\epsfxsize 3 truein \epsfbox {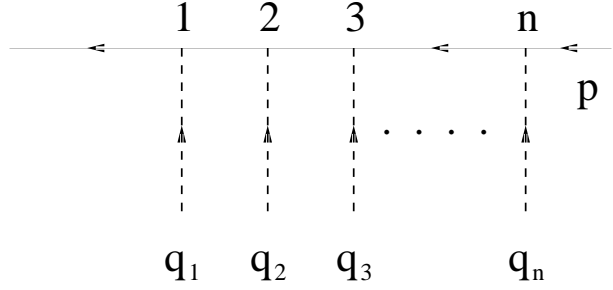}}
\nobreak
\vskip -2 cm\nobreak
\vskip -.5 cm
\caption{A tree diagram, to be designated by the order of the dotted
lines from left to right, as
$[12\:n]$.}
\end{figure}

A tree will be denoted according to the ordering of lines $q_i$, so
the one shown in Fig.~7 is then $[12\:n]$. Its scalar tree amplitude is
\begin{eqnarray}a[12\:n]=-2\pi i \delta(\sum_{j=1}^n\omega_j)
\(\prod_{i=1}^{n-1}{1\over \sum_{j=1}^i\omega_j+i\epsilon}\)\ .
\end{eqnarray}
In this subsection we will show that certain sums of amplitudes
of this type can be expressed as a factorized product.

We need some notations.
If $[T_i]$ are tree diagrams, then $[T_1T_2\:T_A]$ is
the tree diagram obtained by merging these $A$ trees together.
For example, if $[T_1]=[123]$ and $[T_2]=[45]$, then $[T_1T_2]=[12345]$.
The notation $\{T_1;T_2;\:;T_A\}$, on the other hand,
is used to denote the {\it set} of all tree diagrams obtained by
{\it interleaving} the trees $T_1, T_2, \:, T_A$ in all possible ways.
This set contains
$(\sum_a n_a)!/\prod_an_a!$ trees if $n_a$ is the number of gluon lines in the
tree $T_a$. In the example above, $\{T_1;T_2\}$
contains the following $5!/3!2!=10$ trees: [12345], [12435], [12453],
[14235], [14253], [14523], [41235], [41253], [41523], and [45123].

Correspondingly,
$a\{T_1;T_2;\:;
T_k\}$ will represent the sum of the amplitudes $a[T]$ for every tree $T$
in this set.
The {\it factorization formula} \cite{1} states that this sum can be
factorized into a product:
\begin{eqnarray}a\{T_1;T_2;\:;T_A\}=\prod_{a=1}^Aa[T_a]\equiv
a[T_1|T_2|\cdots|T_A]\ .
\end{eqnarray}

\n\underbar{\it Proof}:\quad Let $[T]=[t_1t_2\:t_n]$
be a tree and $a[T]$ its scalar amplitude defined by (3.2). Then
\begin{eqnarray}a[T]
&=&(-i)^n\int_{T_o}d^n\tau\exp\(i\sum_{i=1}^n\omega_{t_i}
\tau_{t_i}\)\nn\\
\int_{T_o}d^n\tau&\equiv&\int_{-\infty}^\infty
d\tau_{t_n}\int_{\tau_{t_n}}^\infty
d\tau_{t_{n-1}} \:\int_{\tau_{t_{2}}}^\infty d\tau_{t_1}\ ,
\end{eqnarray}
where the integration region $T_o$ is
defined by the ordering $\infty> \tau_{t_1}\ge \tau_{t_2} \ge \cdots \ge
\tau_{t_n}> -\infty$.  When summed
over all $T\in\{T_1;T_2; \:\}$, the integration variables $\tau_{t_a}$
retains only the ordering within each individual tree $T_i$, and for each
tree they integrate from $-\infty$ to $+\infty$. Using (3.4) again for
individual trees $T_i$, we obtain (3.3). \quad {\it (end of proof)}
\bigskip

There is a close affinity between the factorization formula and the
string-like representation. See Ref.~[12] for a discussion on this point.

The right-hand-side of (3.3) can be regarded as the  amplitude
$a[T_1|T_2|\:|T_A]$ of the {\it cut diagram} $[T_1|T_2|\:|T_A]$.
This is just the diagram $[T_1T_2\:T_A]$ with cuts put on the propagators
right after $T_1, T_2,\:,T_{A-1}$.
The {\it cut amplitude} $a[T_1|T_2|\:|T_A]$ is obtained from
$a[T_1T_2\:T_A]$ by using the Cutkosky propagator
$-2\pi i\delta(\sum_j\omega_j)$ on the cut lines instead of
the Feynman propagator $(\sum_j\omega_j+i\epsilon)^{-1}$.

For example, the cut diagram $[241|3|65]$ is shown in
Fig.~8 and the corresponding
cut amplitude is $a[241|3|65]=a[241]a[3]a[65]$. Note that the sections
between cuts in a cut amplitude can be permuted at will, so for example,
$a[241|3|65]=a[3|65|241]$. We shall refer to this as the {\it commutative
property} of the abelian cut amplitude.

\begin{figure}
\vskip -0 cm
\centerline{\epsfxsize 3 truein \epsfbox {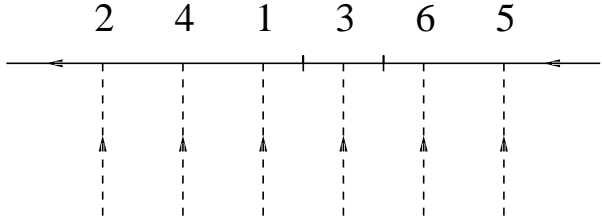}}
\nobreak
\vskip -12 cm\nobreak
\vskip 1.5 cm
\caption{The abelian cut diagram $[241|3|65]$.}
\end{figure}

The factorization formula can be used to sum up {\it loop}
amplitudes as well. This is because
(3.3) is valid whether $q_i$ are onshell or not, so the tree in Fig.~7 or 8
may very well be just a section of a much larger loop diagram, in which case
(3.3) gives a sum of loop diagrams in terms of a
single cut diagram. We shall call these {\it abelian cut diagrams}.

Abelian cut diagrams resemble Cutkosky cut diagrams but they are quite
different.
Cuts here are only put on a high-momentum tree, and the resulting
cut diagram represents a summation of Feynman diagrams rather than
the discontinuity of one of them.

In the simple case when every tree $[T_i]=[i]$ is a vertex containing
a single line, the factorization formula can be written as
\begin{eqnarray}a[1;2;\:;A]=a[1|2|\:|A]\ .
\end{eqnarray}
This is called the {\it eikonal formula} and it has been known for a long
time \cite{7}.

A single abelian cut diagram represents a sum of Feynman diagrams,
nevertheless it is easier to evaluate a single cut diagram than even a single
Feynman diagram.
This is mainly because there are fewer flow diagrams present in a cut diagram
than a Feynman diagram.
To see that, suppose cuts are made on the tree of the top line, which is what
we will usually do.
A cut propagator
is given by $-2\pi i\delta(sX)$ (see Sec.~2.2.2 for notation). Now $\rs X$
is the \- flow through the line, which must now be zero because of the presence
of the $\delta$-function,
so a cut propagator is `cut'
also in the sense of severing the flow. This limits the various ways
the \- flow can go around the diagram and hence the number of possible
flow diagrams.

\subsection{The Sum of QED $s$-Channel-Ladder Diagrams}

A $(2n)$th order $s$-channel-ladder diagram for electron-electron scattering
is obtained by tying together
the $n$ photons from two electron trees like Fig.~7.
There are all together $n!$ diagrams of this type depending on
how the lines from the two trees are tied. Fig.~9 shows
the 6 diagrams for the case $n=3$.

\begin{figure}
\vskip -0 cm
\centerline{\epsfxsize 3 truein \epsfbox {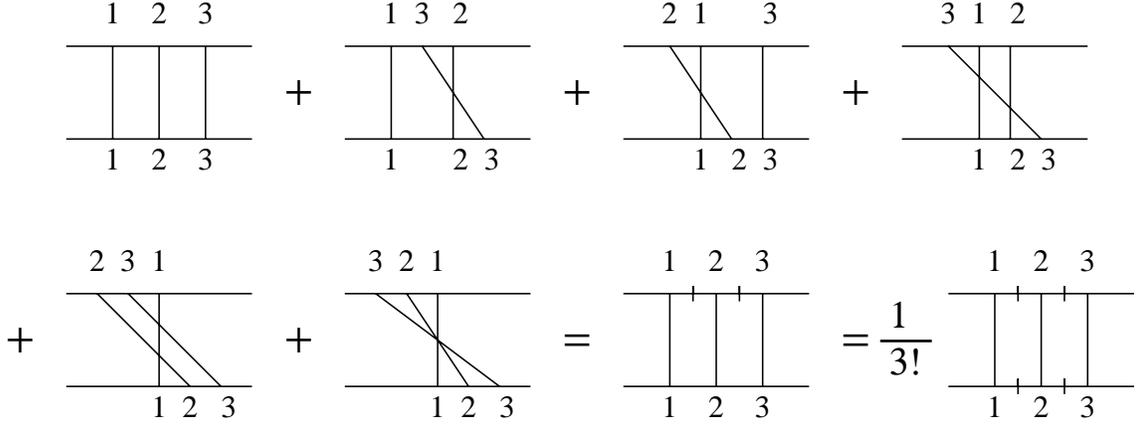}}
\nobreak
\vskip -10 cm\nobreak
\vskip 1.5 cm
\caption{The six $s$-channel-ladder diagrams for $n=3$.}
\end{figure}

These $n!$ diagrams can be labelled by the order of the photons
$[\sigma_1\sigma_2\:\sigma_n]$
on the upper tree, when their order on the lower tree is fixed to be $[12\:n]$.

Use (3.5) to sum over the $n!$ permuted upper trees,
the result is an upper tree with all its propagators cut, as shown
in the diagram between the two equal signs in Fig.~9 for
the case $n=3$. We may now use
the commutative
property of an abelian cut amplitude to symmetrize the photon lines attached
to the {\it lower tree}. Use (3.5) again but now on the lower tree,
we get a double-cut planar diagram, where all the electron propagators on the
upper
and on the lower trees are cut. A factor $n!$ is needed
to convert the
symmetrization into addition.
For $n=3$ this is shown in the last diagram of Fig.~9.

The double-cut diagram is now easy to compute and one gets
\begin{eqnarray}
{T_n\over 2s}&=&-{1\over n!}\int \[\prod_{i=1}^{n-1}(2s)g^2Dq_{i\perp}{1\over
a_i} {i\over 8\pi^2}dq_{i+}dq_{i-}(-2\pi i)^2\delta(\rs q_{i+})\delta
(\rs q_{i-})\]{g^2\over a_n}\nonumber\\
&=&-{g^{2n}\over n!}(-i)^{n-1}I_n(\Delta)\ ,
\end{eqnarray}
where $I_n$ is defined in (2.13) and (2.15).

Let us compare the result with those obtained in (2.19) by direct
calculations. For $n=1$, this should equal to $A$ of (2.19) and indeed it does.
For $n=2$, it should be
$B1+B2$ and again it is.
Note that individually $B1$ and
$B2$ are of order $\ln s$. To compute the sum which is of $O(1)$
correctly we need to know each {\it Feynman diagram} to subleading-log
accuracy. A {\it cut diagram} is already a sum so
each can be computed just in
the leading-log approximation as was done in (3.6).
This ability to avoid subleading-log calculation is the main advantage of
the cut diagrams.

This is further illustrated by looking at $n=3$, whose correct result
is shown in (3.6) to be $g^6I_3/3!$. This cannot even be obtained from (2.19)
because the calculation of order $g^6$ there is accurate only to leading log.

In the impact-parameter space, (3.6) is given by
\begin{eqnarray}{\tilde T_n(b)\over 2s}&=&-i{1\over n!}\[-ig^2\tilde
I_1(b)\]^n\ . \end{eqnarray}
Summing over all $n$, we get an impact-parameter representation of the
$T$-matrix in the eikonal form:
\begin{eqnarray}
{T\over 2s}&=&\sum_{n=1}^\infty \int d^2b \exp(iq_\perp\.b)
{\tilde T_n(b)\over 2s}\nonumber\\
&=&-i\int d^2be^{iq_\perp\.b}\{\exp[-ig^2\tilde I_1(b)]-1\}\ .
\end{eqnarray}
This elegant expression has been known for a long time \cite{6}.
It is call the eikonal formula for high-energy scattering
and is where the `eikonal formula' (3.5) derived its name from.
This $S$-matrix is unitary, with $-g^2\tilde I_1(b)$ being essentially
the phase shift an electron accumulated when it shoots by another electron
at an impact parameter $b$. High energy approximation enters by assuming
the electrons travel essentially at a straight-line trajectory with fixed
impact parameter (perpendicular distance between the scattered electrons)
fixed.
It leads to a total cross-section independent
of energy. If pair productions in the intermediate states are included,
then a rising total cross-section can be obtained \cite{6}.

\subsection{Other Sum Rules for QED}

In the last subsection we illustrated the power of the eikonal formula
(3.5) in computing sums of all ladder diagrams. We shall give two examples in
this subsection
to illustrate the more general factorization formula (3.3) in computing other
sums. These two examples are shown in Figs.~10 and 11.

\begin{figure}
\vskip -0 cm
\centerline{\epsfxsize 3 truein \epsfbox {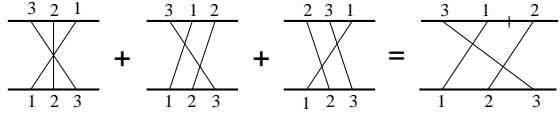}}
\nobreak
\vskip 1 cm\nobreak
\vskip -.5 cm
\caption{An example of using the factorization formula (3.3) to convert
a sum of Feynman diagrams into a single abelian cut diagram.}
\end{figure}

\begin{figure}
\vskip -0 cm
\centerline{\epsfxsize 3 truein \epsfbox {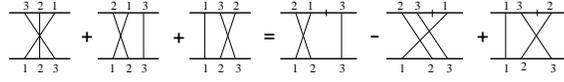}}
\nobreak
\vskip 1 cm\nobreak
\vskip -.5 cm
\caption{Another example of using the factorization formula (3.3) to convert
a sum of Feynman diagrams into a sum of abelian cut diagrams.}
\end{figure}

In each case, (3.3) is applied to
the upper tree with the lower tree held fixed. The two cases
correspond respectively to the identities
\begin{eqnarray}
a\{31;2\}&=&a[31|2]\\
a[321]+a[231]+a[132]&=&a\{21;3\}-a\{23;1\}+a\{13;2\}\nonumber\\
&=&a[21|3]-a[23|1]+a[13|2]\ .
\end{eqnarray}

We shall now compute the abelian cut diagrams and compare them
with the result obtained by computing Feynman diagrams.

The flow diagrams for the four abelian cut diagrams
are shown in Fig.~12. The flow in each case is unique, and
the two poles for the two \+ integrations can all be taken at the
bottom electron line. The computation is as before, and the result for
Figs.~12 (a),(b),(c),(d) are respectively

\begin{figure}
\vskip -0 cm
\centerline{\epsfxsize 3 truein \epsfbox {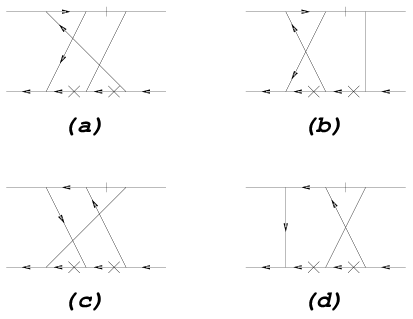}}
\nobreak
\vskip -10 cm\nobreak
\vskip 1.5 cm
\caption{Flow diagrams to compute the right-hand-side of eq.~(3.9)
(diagram (a)) and eq.~(3.10) (diagrams (b),(c),(d)).}
\end{figure}

\begin{eqnarray}
T_a&=&-g^6(2s)(-i){1\over 2\pi}(-ln s)I_3(\Delta)=-T_d\nonumber\\
T_b&=&T_c=0\ .
\end{eqnarray}
The reason for the second equation is because in both 12(c) and 12(d)
the arrows in one of the two loops are all in the same directions.

The sum rules displayed in (3.9), (3.10), Fig.~10, and Fig.~11 demand that
\begin{eqnarray}
C16+C19+C20&=&{T_a\over 2s}\\
C16+C18+C17&=&{1\over 2s}(T_b-T_c+T_d)\ ,
\end{eqnarray}
which can be seen to be true from (2.19), (3.12), and (3.13).

In these two sum rules the leading $\ln s$
is not cancelled, but the complicated
transverse function $J_3$ in (2.19) is.

\begin{center}
\section{NONABELIAN CUT DIAGRAMS}
\end{center}
\setcounter{equation}{0}
\bigskip
\subsection{Multiple Commutator Formula}
\medskip

In this section we consider the generalization of the eikonal formula
(3.5) to nonabelian amplitudes. In other words,
given a tree diagram (Fig.~7) with nonabelian vertices $t_a$ and amplitude
$a[12\:n]\.t[12\:n]\equiv a[12\:n]\.t_1t_2\:t_n$, we wish to find an
expression for the sum of the $n!$ permuted amplitudes
to be expressed in terms of cut diagrams. This turns out to be
the {\it multiple commutator formula} \cite{1}, which states that
\begin{eqnarray}\sum_{\sigma\in S_n}a[\sigma]t[\sigma]=\sum_{\sigma\in S_n}a[\sigma_c]
t[\sigma_c']\ ,
\end{eqnarray}
where $S_n=\{1;2;\:;n\}$, $a[\sigma]_c$ is an abelian cut amplitude (see
Sec.~3.1) for the cut diagram $[\sigma]_c$,
and $t[\sigma]'_c$ is the corresponding nonabelian
factor computed from the {\it complementary cut diagram} $[\sigma]'_c$.

The cut diagram $[\sigma]_c$ is obtained by putting cuts in the
Feynman tree $[\sigma]$ in the following way.
Proceeding from left to right, a cut is put after
a number iff there is not a smaller number to its right.
An external line is considered as a cut so there is never any need
to have an explicit cut at the end of a tree.

Here are some examples
of where cuts should be put:
$[1234]_c=[1|2|3|4]$, $[3241]_c=[3241]$, and $[2134]_c=[21|3|4]$.

The {\it complementary cut
diagram} $[\sigma'_c]$ is one where lines cut in $[\sigma_c]$ are not cut in
$[\sigma'_c]$, and vice versa. Thus $[1234]'_c=[1234]$,
$[3241]'_c=[3|2|4|1]$, and $[2134]'_c=[2|134]$.

The nonabelian factor
$t[\sigma'_c]$ is  obtained from $t[\sigma]$ by replacing the product of
nonabelian matrices separated by cuts with their commutators. For example,
$t[1234]'_c=t[1234]=t_1t_2t_3t_4$,
$t[3214]'_c=t[3|2|4|1]=[t_3,[t_2,[t_4,t_1]]]$,
and $t[2134]'_c=t[2|134]=[t_2,t_1]t_3t_4$.

Let me now illustrate this by writing down explicitly the multiple
commutator formula for $n=3$.
In that case $S_3=\{1;2;3\}$ contains 6 trees, and
the right-hand side of (4.1) is given by
\begin{eqnarray}
&a[1|2|3]t[123]+a[1|32]t[13|2]+a[21|3]t[2|13]+a[231]t[2|3|1]&\cr
&+a[31|2]t[3|12]+a[321]t[3|2|1]\ .
\end{eqnarray}

As before, the trees in $S_n$ can be imbedded in much bigger Feynman
diagrams, in which case the right-hand side of (4.1) and the corresponding
{\it nonabelian cut diagrams} can be used to sum up loop diagrams as well.

Application of this formula will be discussed in Secs.~5 and 6. Before
doing that, we shall sketch the proof of (4.1) in the next subsection.

\subsection{Folding Formula}

The proof of the multiple commutator formula relies on another combinatorial
formula which I call the
{\it folding formula} \cite{1}:
\begin{eqnarray}a[RoS]=\sum_{k=0}^N(-)^k
a\{R;
\tilde \sigma_{1,k}.o\}a[\sigma_{k+1,N}]\equiv
\sum_{k=0}^N(-)^k
a\{R;
\tilde \sigma_{1,k}.o|\sigma_{k+1,N}\}\ .
\end{eqnarray}
Let me explain the notation and the meaning of this formula.

Consider the amplitude of a tree $[RoS]$, put together from a tree $[R]$,
followed by a line $[o]$, and then followed by another tree $[S]$.
The tree $[S]$ is assumed to have $N$ lines,
 with $[\sigma_{1,k}]$ denoting its subtree formed from the first $k$ lines,
and $\sigma_{k+1,N}$ its subtree formed from the remaining $N-k$ lines.
In other words, $[S]=[\sigma_{1,k}\sigma_{k+1,N}]$ for every $k$.
The trees $\sigma_{1,0}$ and $\sigma_{N+1,N}$ are taken to be the null tree
$[\emptyset]$. By definition, $a[\emptyset]=1$. The notation
$\tilde\sigma_{1,k}$ means the tree $\sigma_{1,k}$ read in the
reverse order. For example,
if $[S]=[14327856]$, then $\sigma_{1,3}=[143]$,
$\sigma_{4,8}=[27856]$, and $\tilde\sigma_{1,3}=[341]$.

Let $[t]$ be a tree, with or without cuts.
The symbol $\{T_1;T_2;\:;T_A.t\}$ is taken to mean
the set of trees $[Tt]$ for all $[T]\in\{T_1;T_2;\:;T_A\}$, and
$a\{T_1;T_2;\:;T_A.t\}$ is the sum of these amplitudes $a[Tt]$.

The folding formula (4.3) shows how $a[RoS]$ can be expressed
as sums of products of amplitudes with line `$o$' moved to the end of
the tree in each case. Alternatively the line `$o$' is moved to a
position just before a cut and by the commutative property of abelian
cut amplitudes (see Sec.~3.1) this is just as good as moving it to the
end.

This formula is
called the `folding formula', or the {\it cutting and folding formula},
because the right-hand side can be obtained from the left-hand side
first by {\it cutting} off the tree
$\sigma_{k+1,N}$ at the end of $[RoS]$, and
then {\it folding} the remaining tree about the point $o$, as shown in
Fig.~13. Finally, both branches of the folded tree should be interleaved
to obtain the trees
$\{R;\tilde\sigma_{1,k}.o\}$ in (4.3).

\begin{figure}
\vskip -0 cm
\centerline{\epsfxsize 3 truein \epsfbox {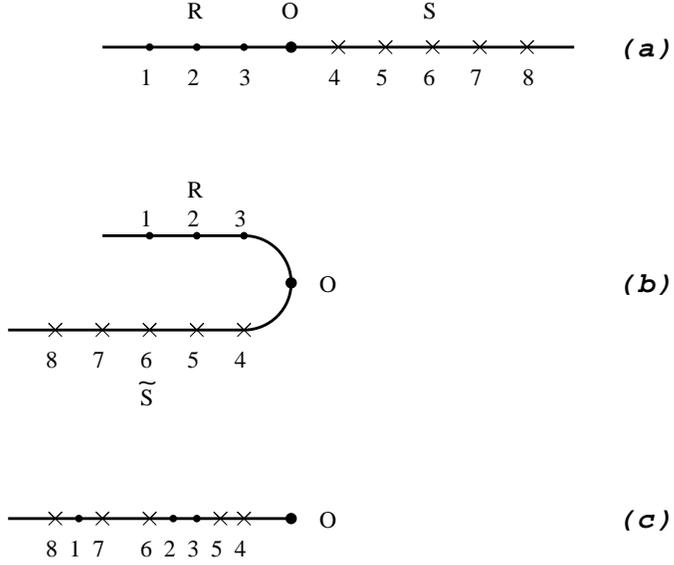}}
\nobreak
\vskip 1 cm\nobreak
\vskip 1.5 cm
\caption{The folding formula illustrated. The cutoff part is not drawn.}
\end{figure}

The folding formula is proven from the factorization formula (3.3)
by an inductive and  combinatorial argument \cite{1}. I shall not repeat
the somewhat involved arguments here, but I will work out some simple
cases to illustrate the formula and its proof.

Suppose we want to move line 1 of
$a[2134]$ to the extreme right. Then in the notation of (4.3), we are faced
with a situation where
$[R]=[2], [o]=[1]$, and $[S]=[34]$. Hence (4.3) gives
\begin{eqnarray}
a[2134]&=&a[21|34]-a\{2;3.1\}a[4]+a\{2;43.1\}\nonumber\\
&=&a[21|34]-a[231|4]-a[321|4]+a[2431]+a[4231]+a[4321]\ .
\end{eqnarray}

Let me now show you how the proof goes for simple cases. Suppose
$[S]=[s]$ is a single line. Then (3.3) implies
$a[Ro|s]=a\{Ro;s\}=a[Ros]+a[R;s.o]$, which leads immediately to the
folding formula $a[Ros]=a[Ro|s]-a\{R;s.o\}.$ If $[S]=[s_1s_2]$, then
the proof goes in a similar way but is more involved. In that case we
can use (3.3) to get $a[Ro|s_1s_2]=a\{Ro;s_1s_2\}=
a[Ros_1s_2]+a\{R;s_1.os_2\}+a\{R;s_1s_2.o\}$. The term
$a\{R;s_1.os_2\}$  can be computed from the $N=1$ formula,
which yields $a\{R;s_1.o\}a[s_2]-a\{R;s_1;s_2.o\}$.
Substituting this back and noticing that
\be a\{R;s_1s_2.o\}-a\{R;s_1;s_2.o\}=-a\{R;s_2s_1.o\}\ ,\ee
we obtain
the desired $N=2$ result
\begin{eqnarray}a[Ros_1s_2]=a[Ro|s_1s_2]-a\{R;s_1.o\}a[s_2]+a\{R;s_2s_1.o\}\ .
\end{eqnarray}

Similar proof goes for any $N$.

I will now show how the multiple commutator formula (4.1) can be obtained
from the folding formula (4.3).

Take any abelian Feynman amplitude $a[\sigma]$. Use (4.3) to cut and to
fold, so as to move the number `1' to the end. The result is a sum of
product of two $a[\cdots]$, the first one having the number `1' at the end,
and the second one not containing the number `1'.
Use (4.3) again to cut and to fold the second $a[\cdots]$,
now to move the smallest number in its
argument to the end,
and so on. The final result is a complicated sum of products of several
$a[\dots]$, each having the smallest number within its argument located
at the end. In other words, each is a cut amplitude $a[\rho]_c$
of some cut diagram $[\rho]_c$ determined in the way specified in Sec.~4.1.

Conversely, given a $[\rho]_c$, we can obtain all the $[\sigma]$'s
that gave rise to it this way by repeated unfoldings and gluings.
To see how this works in more detail, let us first consider the case when
$[\rho]_c$ contains no cut. That means the number `1' is at the end,
so it is of the form $[\rho]_c=[\tau 1]$. To
unfold the tree $[\tau 1]$, pick out any two interleaving subsets $[\tau_1]$
and $[\tilde\tau_2]$ from $[\tau]$, in all possible ways, {\it viz}.,
find all such $\tau_i$ so that $[\tau 1]\in\{\tau_1;\tilde\tau_2 .1\}$. Then
the desired trees after unfolding  are
$[\sigma]=[\tau_1 1\tau_2]$. The sign involved in (4.3) is
$(-)^k$,
where $k$ is the number of boson lines in tree $[\tau_2]$.

The nonabelian factors associated with $a[\sigma]$ is $t[\sigma]$, hence the
nonabelian factor associated with
$a[\rho]_c=a[\tau 1]$ is $\sum (-)^kt[\tau_1 1\tau_2]$,
which is nothing but the multiple commutator $t[\rho]'_c$.

We have now considered the case when $[\rho]_c$ has no explicit cuts.
If $[\rho]_c$ has explicit cuts, then we can apply the argument above
to each of the cut sections, to obtain a multiple commutator for
the nonabelian factor of each cut section. Thus the nonabelian factor
for $a[\rho]_c$ is always $t[\sigma]'_c$ as specified in Sec.~4.1,
and the multiple commutator formula is proved.
\quad ({\it end of proof})

\begin{center}
\section{QUARK-QUARK SCATTERING TO $O(g^6)$}
\end{center}
\setcounter{equation}{0}
\bigskip
\subsection{Color Factors}

The color factor of a quark-gluon vertex is $t_a$, and that for a triple
gluon vertex is $if_{abc}$. Putting these together, and using (1.8) and
(1.9), or graphically Fig.~1,
the color factor for a QCD Feynman diagram can be computed \cite{6}. In what
follows we shall concentrate on `quark-quark' scattering where
the `quark' is allowed to carry any color in an $SU(N_c)$ theory
with any $N_c$. In other words, the `quarks' below could very well have
been gluons as far as color is concerned.

Fig.~14 contains an illustration as to how this can be done. For
other examples see Ref.~[6].

\begin{figure}
\vskip -0 cm
\centerline{\epsfxsize 3 truein \epsfbox {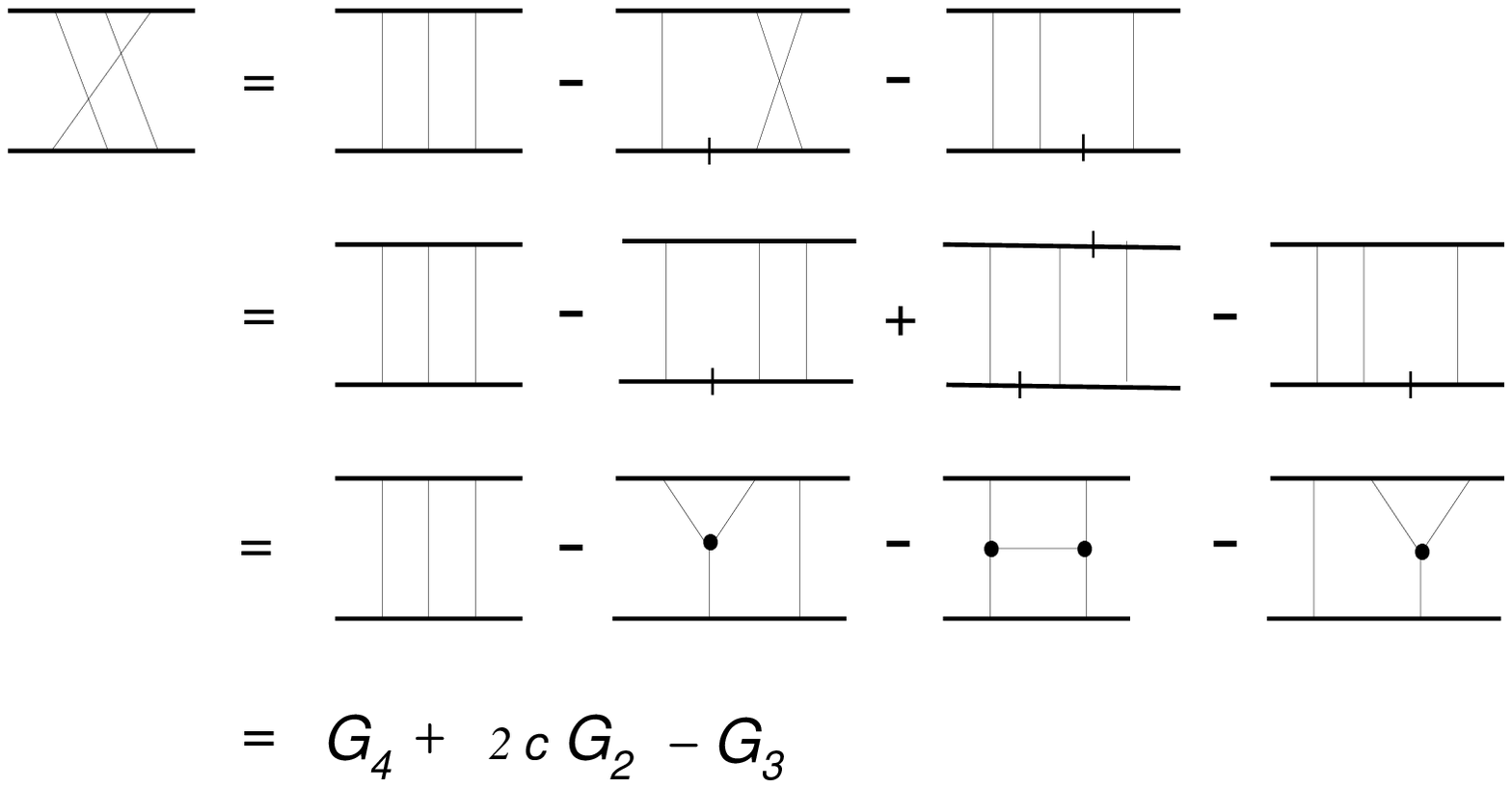}}
\nobreak
\vskip -2 cm\nobreak
\vskip 1.5 cm
\caption{An illustration for the computation of the color factor of a Feynman
diagram}
\end{figure}

The color factor for a nonabelian cut diagram is computed in a similar
way. Fig.~15 contains one such example.

\begin{figure}
\vskip -1 cm
\centerline{\epsfxsize 3 truein \epsfbox {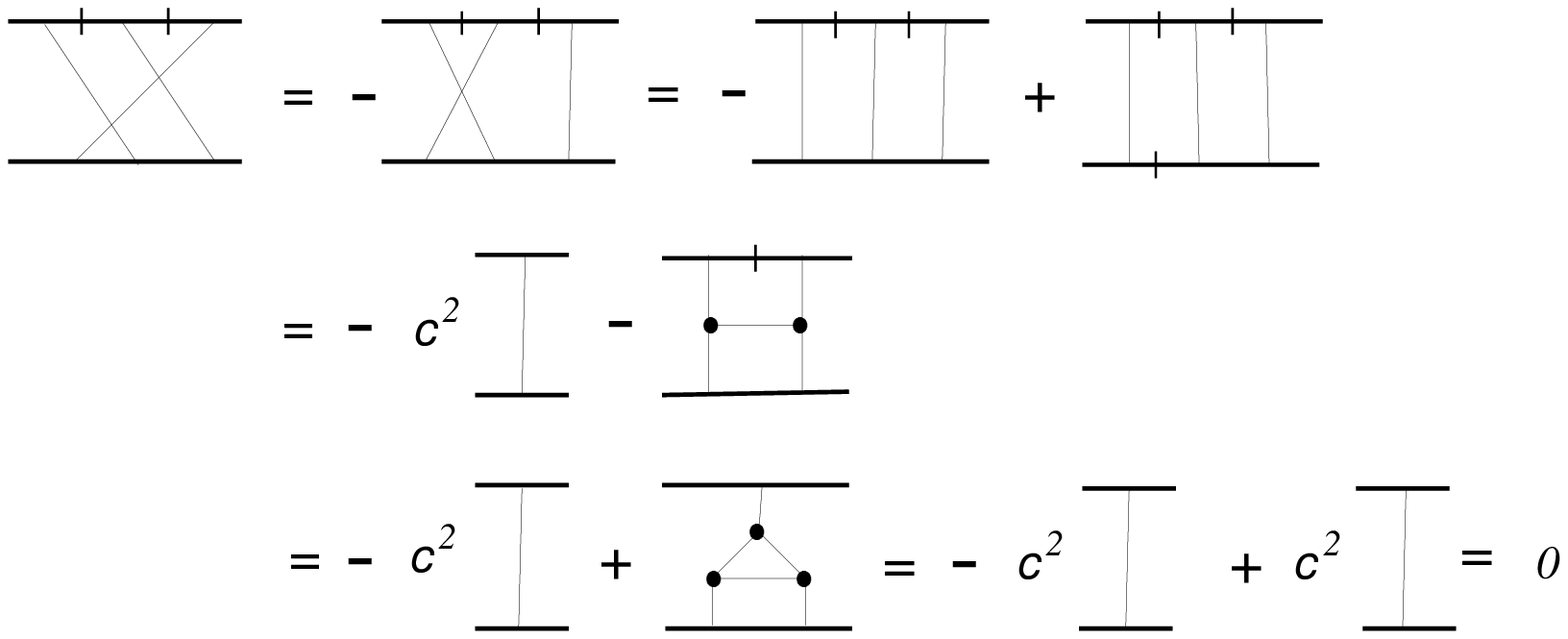}}
\nobreak
\vskip -1 cm\nobreak
\vskip 1.5 cm
\caption{An illustration for the computation of the color factor of a
nonabelian cut diagram.}
\end{figure}

To order $O(g^6)$, the color factors can all be reduced by this method to
combinations of the four shown in Fig.~16 \cite{2}.

\begin{figure}
\vskip -4 cm
\centerline{\epsfxsize 3 truein \epsfbox {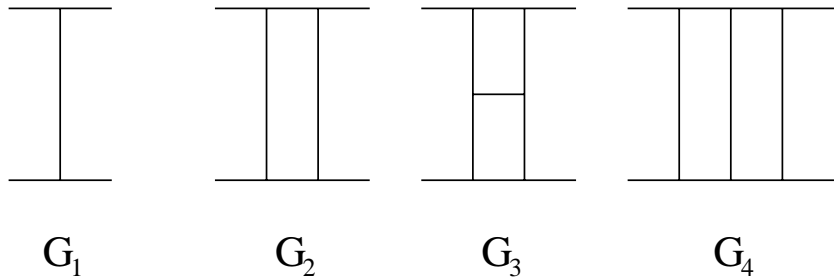}}
\nobreak
\vskip -9 cm\nobreak
\vskip 1.5 cm
\caption{The four color factors for $O(g^6)$ diagrams.}\end{figure}

\subsection{Sum of Feynman Diagrams}

To $O(g^6)$ the QCD Feynman diagrams for quark-quark scattering are
shown in Fig.~6. Each of their color factors can be computed
by the method of the last subsection, and each of its spacetime amplitude
can be computed as in Sec.~2.2. In fact, those diagrams common to QED have
already been listed already in (2.19). For others, see Ref.~[6] for details
of the computations.

Putting these together, The contribution from each diagram in Fig.~6
to $T/(2s)$ is ($\beta=g^2/2\pi$)
\begin{eqnarray}
A&=&-g^2I_1\!\cdot\!{\bf G}_1\nonumber \\
B1&=&- g^2\beta\ln\left(  se^{-\pi i}\right) I_2\!\cdot\!{\bf G}_2\nonumber \\
B2&=&+g^2\beta(\ln s) I_2\!\cdot\!({\bf G}_2+c{\bf G}_1)\nonumber \\
\overline{C1}&=&+g^2\beta^2\ln^2\left(se^{-\pi i}\right)\left[ {1\over 2}
\Delta^2I_2^2-J_2I_2\right] \!\cdot\!{\bf G}_3\nonumber \\
\overline{C2}&=&-g^2\beta^2({\ln^2s})\left[ {1\over 2}
\Delta^2I_2^2-J_2I_2\right] \!\cdot\!({\bf G}_3+c^2{\bf G}_1)\nonumber \\
C3&=&+g^2\beta^2\{\ln^2\left(se^{-\pi i}\right)-{\ln^2s}\}
{1\over 4}J_2I_2\!\cdot\!({\bf G}_3-c{\bf G}_2)=C4=C5=C6\nonumber \\
C7&=&-g^2\beta^2\ln^2\left(se^{-\pi i}\right)
{1\over 4}J_2I_2\!\cdot\!(-c{\bf G}_2\
)=C8=C9=C10\nonumber \\
C11&=&+g^2\beta^2({\ln^2s}){1\over
4}J_2I_2\!\cdot\!(-c{\bf G}_2-c^2{\bf G}_1)=C12=C13=C14\nonumber \\
C15&=&-g^2\beta^2(\ln s) 2J_3\!\cdot\!{\bf G}_4\nonumber \\
C16&=&-g^2\beta^2(\ln s) 2J_3\!\cdot\!({\bf G}_4-{\bf G}_3+3c{\bf G}_2+c^2
{\bf G}_1)\nonumber \\
C17&=&+g^2\beta^2(\ln s)(J_3+\pi iI_3)\!\cdot\!({\bf G}_4+c{\bf G}_2)=C18\nonumber \\
C19&=&+g^2\beta^2(\ln s)(J_3-\pi
iI_3)\!\cdot\!({\bf G}_4-{\bf G}_3+2c{\bf G}_2)=C20\
\end{eqnarray}
The functions $J_3$ and $I_n$ were already given in  (2.21) and (2.16).
The function $J_2$ is
\begin{eqnarray}J_2(\Delta)=\int{d^2q_\perp\over(2\pi)^2}{1\over q_\perp^2}\ .
\end{eqnarray}

The sum of these diagrams is
\begin{eqnarray}
{1\over 2s}T&=&-g^2I_1\left[ 1-
\overline{\alpha} \ln s+{1\over 2!}\overline{\alpha} ^2
{\ln^2s}\right] \!\cdot\!{\bf G}_1
+{i\over 2} (g^4I_2-{c\over\pi}g^6 I_3\ln s)\!\cdot\!{\bf G}_2\nonumber \\
&+&{i\over 2\pi}\ln s\left[ g^6I_3-{1\over 2}g^6\Delta^2I_2^2\right]
\!\cdot\! {\bf G}_3
+{1\over 3!}g^6I_3\!\cdot\! {\bf G}_4\ ,
\end{eqnarray}
where
\begin{eqnarray}\overline{\alpha} (\Delta)\equiv {c\over 2\pi}g^2
\Delta^2I_2(\Delta)\ .
\end{eqnarray}

There are several things worth noting:

\begin{enumerate}
\item
In the fourth order, the leading term proportional to $(\ln s)$ is cancelled
out between $B1$ and $B2$ in the color amplitude ${\bf G}_2$,
though not in ${\bf G}_1$.
\item In the sixth order, the leading $(\ln s)$ contributions to ${\bf G}_4$
from $C15$ to $C20$ also add up to zero. The expressions given in (5.1)
are not accurate enough to deal with the subleading terms as already noted
in Sec.~3.2.
The term in (5.3) proportional to ${\bf G}_4$ is obtained separately from
the eikonal formula.
\item {\it As a result of these cancellations},
the coupling-constant and energy dependences of the ${\bf G}_i$
amplitude is of the form $g^{2m}(g^2\ln s)^p$, where
$m=1,2,2,3$ respectively for the color factors $i=1,2,3,4$, and $p$ is
determined by the order of the perturbation. Referring to Fig.~16, we see that
$m$ is simply the number of vertical lines
in the color factor, or as we shall see in the next section, the number of
reggeons being exchanged. It is important
to note that this dependence is not true for individual Feynman diagrams.
\item To leading-log accuracy, the functions $J_2$ and $J_3$ get
cancelled out. Only $I_n$ appears in the final result.
\item The quantity in the square bracket in the ${\bf G}_1$
term of (5.3) is simply the the first three terms in the expansion of
$\exp(-\overline\alpha\ln s)$. If we define the
(transverse) reggeon propagator to be
\begin{eqnarray}R_1(\Delta,s)=I_1(\Delta)
\exp(-\overline{\alpha} (\Delta)\ln s)\ ,
\end{eqnarray}
and analogous to $I_n=(*I_1)^n$ we define the $n$-reggeon propagator
to be
\begin{eqnarray}R_n(s,\Delta)=(*R_1)^n(s,\Delta)\ ,\end{eqnarray}
then the sum (5.3), to $O(g^6)$, is equal to
\begin{eqnarray}{1\over 2s}T=
-g^2R_1(\Delta,s)\!\cdot\!{\bf G}_1+ {ig^4\over 2!}\left[
R_2(\Delta,s)\!\cdot\! {\bf G}_2+R_{2,1}(\Delta,s)\!\cdot\!{\bf G}_3\right]
+{g^6\over 3!}R_3(\Delta,s)\!\cdot\!{\bf G}_4\ .
\end{eqnarray}
The function $R_{2,1}$ is equal to
\begin{eqnarray}R_{2,1}(s,\Delta)={1\over\pi}
g^2(\ln s)\[I_3-\h \Delta^2I_2^2\]\ .
\end{eqnarray}
The physical interpretation of these equations will be deferred till Sec.~6.1.

\end{enumerate}

\subsection{Sum of Nonabelian Cut Diagrams}

The QCD nonabelian cut diagrams to $O(g^6)$ are shown in Fig.~17,
with a subscript `$c$' to emphasize that they are cut diagrams
and not Feynman diagrams. Their computations can be found in Ref.~[2].

\begin{figure}
\vskip -0 cm
\centerline{\epsfxsize 3 truein \epsfbox {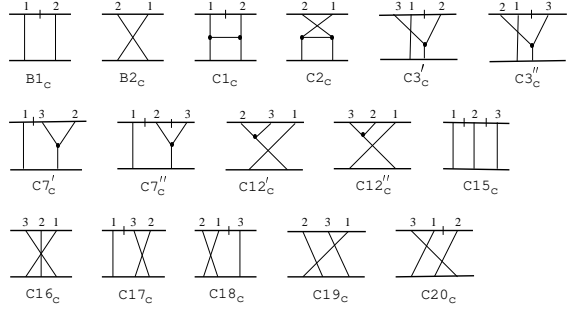}}
\nobreak
\vskip -2 cm\nobreak
\vskip 1.5 cm
\caption{Nonabelian cut diagrams for QCD to order $g^6$.}
\end{figure}

For the $s$-channel-ladder cut diagrams,
the computation is straight forward and one gets their contributions
to $T/(2s)$ to be
\begin{eqnarray}
B1_c&=&+{1\over 2}ig^4I_2\!\cdot\!{\bf G}_2\nonumber \\
B2_c&=&+{c\over 2\pi}(\ln s)g^4 I_2\!\cdot\!c{\bf G}_1\nonumber \\
C15_c&=&+{1\over 6}g^6I_3\!\cdot\!{\bf G}_4\nonumber \\
C16_c&=&-{c^2\over 2\pi^2}(\ln s) g^6J_3\!\cdot\!c^2{\bf G}_1\nonumber \\
C17_c&=&C18_c=0\nonumber \\
C19_c&=&0\nonumber \\
C20_c&=&-{1\over 2\pi} i(\ln s)g^6 I_3\!\cdot\!(c{\bf G}_2-{\bf G}_3)\ .
\end{eqnarray}

Similarly, we choose to cut the line of the planar diagram C1 to obtain
\begin{eqnarray}
\overline{C1}_c&=&-g^4\beta i(\ln s)\left[ {1\over 2}
\Delta^2I_2^2-J_2I_2\right] \!\cdot\!{\bf G}_3\nonumber \\
\overline{C2}_c&=&-g^2\beta^2({\ln^2s})\left[ {1\over 2}
\Delta^2I_2^2-J_2I_2\right] \!\cdot\!c^2{\bf G}_1\ .
\end{eqnarray}

The twelve diagrams C3 to C20 can be divided into four groups of three, each
giving identical contributions, so it is necessary to consider only one
of these groups. The group of C3, C7, C12 have been chosen for that purpose.
There is a symmetry between gluon lines 2 and 3 so we may double this
group and consider it as a sum of six Feynman diagrams.
By applying the multiple commutator formula,
the six cut diagrams shown in Fig.~17 are obtained. Their values are
\begin{eqnarray}
C3_c&=&{1\over 2}(C3_c'+C3_c'')=-g^4\beta i(\ln s){1\over 4}J_2I_2\!\cdot\!{\bf
G}_3\nonumber \\ C7_c&=&{1\over 2}(C7_c'+C7_c'')=0\nonumber \\
C12_c&=&{1\over 2}(C12_c'+C12_c'')=g^2\beta^2(\ln^2\!s){1\over
4}J_2I_2\!\cdot\!(-c^2{\bf G}_1)\ . \end{eqnarray}
It is easy to verify that
they sum up once again to (5.3) as they should.

Comparing this calculation of the cut diagrams with the earlier
calculation of Feynman diagrams, one notes that

\begin{enumerate}
\item The coupling-constant and energy dependence $g^{2m}(g^2\ln s)^p$,
noted previously to be true for {\it sums of Feynman diagrams}
under item `3' of the last subsection, is now true
for {\it individual} nonabelian cut diagrams. In other words,
delicate cancellations of the kind found in Feynman diagrams when they
are summed never appear in cut diagrams. This is one of the main advantages
of dealing with nonabelian cut diagrams.
\item In leading-log approximation, functions $J_2$ and $J_3$ that
appeared in individual Feynman diagrams but get cancelled in the sum
never even occur here.
\end{enumerate}

\bigskip\bigskip
\section{MULTIPLE REGGEONS AND QCD}
\setcounter{equation}{0}
\subsection{The Reggeized Factorization Hypothesis}

Eq.~(5.7) shows that QCD scattering takes on a simple form
at high energies, at least up to $O(g^6)$.
The color factors present are those of Fig.~16, or equivalently
Figs.~18(a), (b), (e), and (c). Interestingly enough, the spacetime amplitudes
are also given by these same figures in Fig.~18,
if they are interpreted as {\it reggeon diagrams}.

\begin{figure}
\vskip -0 cm
\centerline{\epsfxsize 3 truein \epsfbox {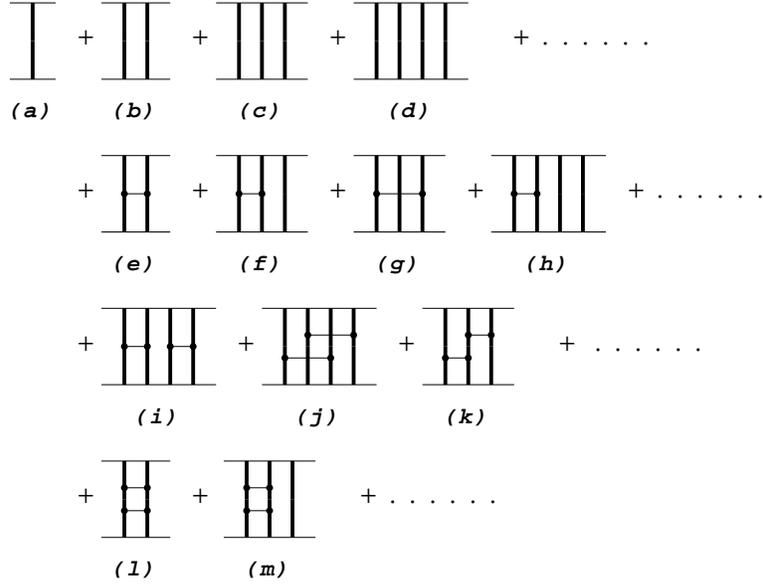}}
\nobreak
\vskip -2 cm\nobreak
\vskip .5 cm
\caption{Reggeon diagrams for quark-quark scattering.}
\end{figure}

The vertical lines
in Fig.~18 are reggeized gluons. The horizontal lines are the
leading particles (top and bottom), or ordinary gluons produced from
the
reggeons. These
are two-dimensional diagrams in the transverse-momentum space:
the (vertical) reggeon propagators are given by $R_1(s,q_\perp)$ of (5.5),
horizontal lines carry no factor, each vertex carries a factor $g$, and loop
integrations for $-T/(2s)$ are given by the measure $-i\int
d^2q_\perp/(2\pi)^2$. The spacetime amplitude for $-T/(2s)$ from an
$m$-reggeized-gluon ($m$rg) diagram without produced gluons (Figs.~18(a), (b),
(c))
is then given by
$(-i)^{m-1}g^{2m}(*R_1)^m(s,\Delta)/m!=(-i)^{m-1}g^{2m}R_m(s,\Delta)/m!$,
where the factor $1/m!$ is a symmetry factor. This agrees with the terms
${\bf G}_1, {\bf G}_2, {\bf G}_4$ in (5.7).

For the reggeon diagram 18(e) with a produced gluon, the rule
is more complicated. A Lipatov vertex $gC_\mu(q_i,q_{i+1})$ is to be placed
at every reggeon-reggeon-gluon junction. For the elastic scattering diagrams
in Fig.~18, they always come in a pair, so the factor is

\be
-2g^2\K(q_i,q_{i+1})&\equiv&
g^2C^\mu(q_i,q_{i+1})C_\mu(\Delta-q_i,\Delta-q_{i+1})\nn\\
&=&-2g^2\[\Delta^2
- {(\Delta-q_{i\perp})^2 q_{i+1\perp}^2 + ( \Delta
 - q_{i+1\perp})^2 q_{i\perp}^2\over (q_{i\perp}-q_{i+1\perp})^2}\]\ ,
\ee
as shown in Fig.~19.

\begin{figure}
\vskip -0 cm
\centerline{\epsfxsize 3 truein \epsfbox {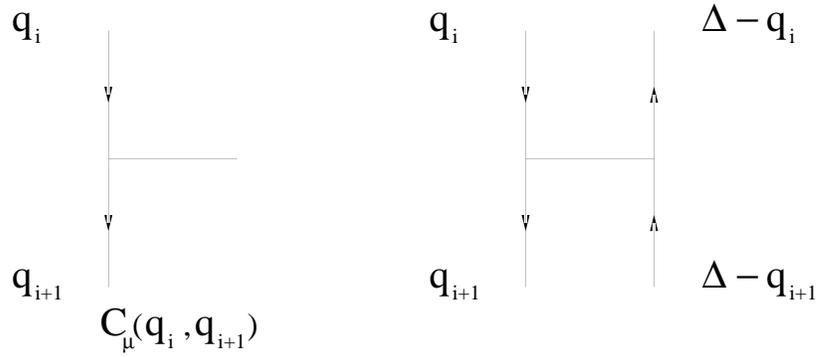}}
\nobreak
\vskip -11 cm\nobreak
\vskip .5 cm
\caption{(a) A Lipatov vertex $gC_\mu(q_i,q_{i+1})$; (b). Two Lipatov
vertices $-2g^2\K(q_i,q_{i_1})$.}
\end{figure}

In addition, the vertical reggeon lines should be
thought of as covering the rapidity range stretched
between the top leading particle and the bottom leading particle.
Each horizontal line connected by
two Lipatov vertices
must be allowed to slide along this whole rapidity range to produce
a factor $(-2\pi i)/2(2\pi)^2\int dy\simeq -i\ln s/4\pi$ per produced gluon. In
$O(g^6)$, the contribution from 18(e) to $-T/(2s)$  is therefore given
by
\be
-{T\over 2s}&=&{1\over 2!}{-i\ln s\over 4\pi}\[{-i\over (2\pi)^2}\]^2g^6\int
d^2q_{1\perp}d^2q_{2\perp} R_1(s,q_{1\perp})
R_1(s,q_{2\perp})R_1(s,\Delta-q_{1\perp})R_1(s,\Delta-q_{2\perp})\.\nn\\
&\.&(-2)\K(q_1,q_2)\nn\\
&\simeq&
{-ig^6\ln s\over 8\pi}\[{-i\over (2\pi)^2}\]^2\int d^2q_{1\perp}d^2q_{2\perp}
{1\over q_{1\perp}^2}  {1\over q_{2\perp}^2}{1\over
(\Delta-q_{1\perp})^2}{1\over
(\Delta-q_{2\perp})^2}\.\nn\\
&&(-2)\[\Delta^2
- {(\Delta-q_{1\perp})^2 q_{2\perp}^2 + ( \Delta
 - q_{2\perp})^2 q_{1\perp}^2\over (q_{1\perp}-q_{2\perp})^2}\]\nn\\
&=&{ig^6\ln s\over 2\pi}\(I_3-\h\Delta^2I_2^2\)\ ,\ee
in agreement with the ${\bf G}_3$ term of (5.7).

In the leading-log approximation where $g^2\ll 1$ and $g^2\ln s=O(1)$,
every $R_i$ and every $\K$ are $O(1)$. Since the produced gluons
must slide along the whole rapidity range mapped out by the reggeon lines,
whenever one of them appears a factor $g^2\ln s=O(1)$ is produced.
Hence an $mrg$ amplitude is of order $g^{2m}$, whether there are produced
gluons or not. This is certainly true in (5.7) but it should also be generally
true for all reggeon diagrams in Fig.~18.

This picture of reggeized exchange is simple and elegant, but
does this $O(g^6)$ result generalize to all orders? In other words,
can the sum of all high-energy elastic scattering Feynman diagrams
be factorized into reggeon diagrams like those in Fig.~18?
If it does, is the reggeon still given by (5.4) to all orders, or something
more complicated is required?
We shall refer to the affirmative answer to both of these questions as the
{\it reggeized factorization hypothesis}.

As discussed above, this hypothesis is completely verified to $O(g^6)$.
It has also been partially checked to $O(g^8)$ and $O(g^{10})$ \cite{8}. The
main focus of this
section is to ask whether this is true to higher orders. To be sure,
we do not know the full answer at the present but some partial answers are
known. For example, using the technique of
nonabelian cut diagrams, we can show that the hypothesis is correct at
least for $s$-channel-ladder diagrams. This will be discussed in Sec.~6.4.
Before doing that, we shall go over briefly what has been known for some time
about multiple reggeon exchanges.

\subsection{BFKL Equation and Unitarity}
By summing $t$-channel-ladder diagrams in the leading-log approximation
one obtains the result that the octet channel is dominated by a reggeized
gluon exchange, with a regge trajectory given by $\alpha(t)=1+\overline
\alpha(\Delta)$, where $\overline\alpha(\Delta)$ is given in (5.4).
The propagator of this reggeized gluon is given by (5.5). \cite{8,9}.

The 2rg (two-reggeized-gluon) exchanges come from
the sum of $t$-channel-reggeon-ladder diagrams
\cite{9}. These are diagrams in which
$n$ ordinary gluons are produced and absorbed by two reggeized gluons.
They are illustrated in Fig.~18 by diagrams (b), (e), and (l) respectively
for $n=0,1,2$.

In the leading-log approximation, the produced gluons are strongly ordered
in rapidity, hence the rungs of the ladder do not cross. A diagram with
$n$ produced gluons has $n+1$ transverse integrations. The integrand
consists of $n$ factors of $\K$ and $2(n+1)$ reggeon propagators $R_1$.
It amplitude has been computed in the following way \cite{9}. First, its
discontinuity is calculated from the Cutkosky rules. Then the whole amplitude
is computed using $s$-channel dispersion relations.

The color-octet amplitude for quark-quark scattering is predominantly real,
and its leading contribution is given by the
1rg (one reggeized gluon exchange) diagram. However, this real part should
also
be obtainable from the imaginary part computed in the 2rg Cutkosky diagram via
a dispersion relation. Explicit calculation shows that this is indeed the case.

The color-singlet amplitude is the Pomeron amplitude originally
proposed by Low and Nussinov \cite{13}. It is
predominantly imaginary,
a factor $g^2$ down from the color-octet amplitude, and dominated by the 2rg
diagrams in the leading-log approximation. Its computation is fairly
difficult
\cite{9}, and it leads to a total cross-section growing like $S^A$, with
$A=g^2(\ln 2)N_c/\pi^2$. This violates the Froissart bound and unitarity,
hence the computation as it stands
must be incomplete. Subleading-log contributions must be added to unitarize
it. These may be the subleading-log contributions
from the same $t$-channel-reggeon-ladder diagrams, or the leading-log
contributions from diagrams associated with more reggeon exchanges, or both.

Subleading-log contributions from Feynman diagrams are very difficult to
compute, so if we must do it that way there would be very little hope
of obtaining any result except in very low orders.
However, if we sum up all multiple reggeon exchange diagrams
as in Fig.~18, then formally $s$-channel unitarity is restored.
From the discussion at the end of the last subsection, we know that an
$mrg$ amplitude is of order $g^{2m}$, so multiple reggeon diagrams
computed to leading-log approximation do contribute to subleading-logs
of the Pomeron amplitude. Thus there is a hope of unitarizing the Pomeron
amplitude by including all these multiple reggeon diagrams computed to
leading-log approximation, provided the reggeized factorization hypothesis is
correct.

\subsection{Nonabelian Cut Diagrams vs Feynman Diagrams}
We saw in the $O(g^6)$ computation of Sec.~5 that delicate cancellations
occur in the sum
 of Feynman diagrams, making it necessary to compute individual Feynman
diagrams to subleading accuracies in order to obtain a finite
sum. In contrast, this never happens in nonabelian cut diagrams, so
they can be computed simply to the leading order, a much easier task.
We shall argue below that this situation is true to all orders provided
the reggeized factorization hypothesis is correct.

Recall that for nonabelian cut diagrams
the color factor is computed from the complementary
cut diagrams. For a diagram to contribute to an $m$rg amplitude, it must
have $m$ uncut propagators on the upper tree, which means that the
corresponding cut diagrams from which the spacetime amplitude is computed must
have $m$ cut lines.

Now an $\ell$-loop amplitude can grow at most like
$(\ln s)^\ell$. However,
each cut propagator with its $\delta$-function
will deprive that loop of a $\ln s$
factor because it replaces integrations like $\int_\epsilon dx/x\sim \ln s$ by
$-2\pi i\int dx\delta(x)\sim 1$.
A diagram with $m$ cuts is thus deprived of at least $m$
factors of $\ln
s$, making it grow at best like $g^(2m)(g^2\ln s)^p$, for a diagram of order
$2(m+p)$. Since $g^2\ln s=O(1)$,
this is of order $g^{2m}$, precisely what an $m$rg amplitude
requires, so it is sufficient to compute each nonabelian cut diagrams just
to the leading order.

In contrast, the corresponding Feynman diagram has no cut in its spacetime part
nor its color part. Its color factor will generally involve $n$rg
diagrams with $n\ge m$. If this diagram contributes to an $n$rg diagram without
delicate cancellation then its spacetime amplitude must be deprived of
more (since $n\ge m$) factors of $\ln s$. On the other hand,
without cuts in the spacetime diagram, there is
no guarantee to have any deprivation of
$\ln s$ at all. The absence of cuts in both therefore conspire to render too
many
$\ln s$ factors compared to those needed for reggeon amplitudes. In order
for the reggeized
factorization hypothesis to be valid, delicate cancellations are then necessary
in the sum to reduce the $\ln s$ powers.

This argument shows that nonabelian cut diagrams are much more suitable for
the computation of sums than Feynman diagram. In fact, there is virtually no
hope of obtaining  sums in higher orders by computing Feynman diagrams
individually.

\subsection{$s$-Channel-Ladder Diagrams}

In this subsection we outline the computation of the $s$-channel ladder
diagrams to show that they satisfy the reggeized factorization hypothesis.
For details of the arguments see Ref.~[10]. For reasons discussed in the
last subsection, we will use nonabelian cut diagrams rather than Feynman
diagrams in our computation.

The Feynman diagrams are
the same diagrams as those computed in Sec.~3.2 for QED to obtain
an eikonal amplitude, leading to a constant total cross-section in that case.
The situation in QCD is quite different. We know from the $O(g^6)$
computation in Sec.~5 that there are $\ln s$ factors present even for
the Pomeron amplitude, so the total cross-section is no longer constant
in QCD.

There is another important difference. The sum of these ladder diagrams
in QED is gauge invariant but that in QCD is not.
To make it gauge invariant we virtually have to include all diagrams,
and that computation is very difficult. What appears below is the
result in the Feynman gauge.

We shall consider quark-quark scattering in which the `quarks' carry
an arbitrary color in an $SU(N_c)$ theory. Since spin is unimportant at
high energies, what is thus obtained is valid for gluon-gluon scattering as
well. The two quarks will be represented by two horizontal lines, and the
$n$ exchanged gluons by vertical or slanted lines as in Fig.~20 for $n=4$.

\begin{figure}
\vskip -0 cm
\centerline{\epsfxsize 3 truein \epsfbox {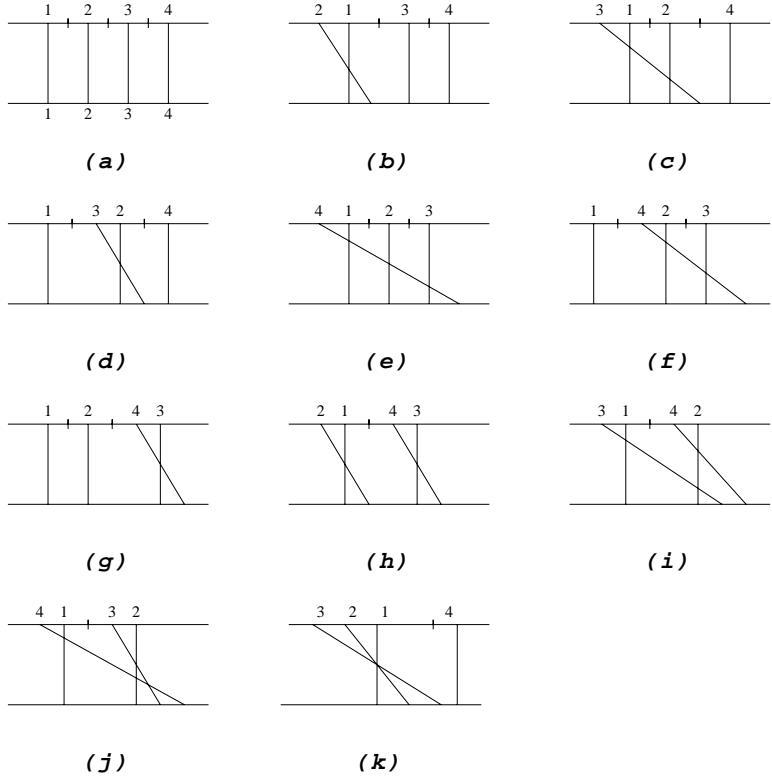}}
\nobreak
\vskip -1.5 cm\nobreak
\vskip .5 cm
\caption{$s$-channel-ladder cut diagrams for $n=4$.}
\end{figure}

We shall choose the {\it planar diagram} to be the one in which all propagators
on the upper line are cut, as shown in Fig.~20(a). This can be accomplished by
numbering the gluons
in the manner shown. For convenience we shall also draw the gluons of the
planar diagram as vertical lines.

All the other nonabelian cut diagrams can be obtained by
pulling the upper ends of a number of gluon lines leftward, in all possible
combinations.
The propagator to the right of a slanted line is uncut, and the propagators to
the right of a vertical line is cut. See Fig.~20. This is so because once
the upper end of a gluon line is moved leftward, there will always be
another gluon with a smaller number to its right.

We shall refer to these $s$-channel-ladder cut diagrams as SC diagrams.
Their complementary diagrams used to compute color factors will be referred
to as SCC diagrams.

As discussed in the last subsection, a nonabelian cut diagram
with $m$ cuts are deprived of at least $m$ would-be $\ln s$ factors.
Such a diagram is said to be {\it saturated} if it is deprived of exactly
$m$ factors of $\ln s$, and no more.
The leading contribution to a $m$rg diagram will come only from the saturated
SC diagrams with $m$ cuts.

By referring to the computations
in Secs.~2.2.3 and 2.2.4, we see that even diagrams with no cuts may not
get its full share of $\ln s$ factors -- Fig.~4 does but Fig.~5 does not.
Like the cases in  Figs.~4 and 5, an SC diagram is saturated iff it does not
have two consecutive
uncut propagators on the upper line. Thus 20(k) is not saturated, but every
other diagram in Fig.~20 is.

We outline now the procedure for computing sums of SC diagrams of any $n$.
For details please consult Ref.~[10].

We shall refer to color factors of the type shown in Fig.~18 as {\it regge
color factors}. Given an SCC diagram, the first step is to use eqs.~(1.8)
and (1.9), or equivalently the graphical rules in Fig.~1, to decompose its
color factor into a sum
of regge color factors. We will call a regge color factor {\it primitive} if
it remains connected after the top and bottom lines are removed.
Thus the regge color factors in Figs.~18(a), 18(e), 18(k), and 18(l) are
primitive, but none of the others explicitly shown in Fig.~20 are.
18(l) and 18(m) do not occur in SCC diagrams so we will ignore them. The other
three primitive ones 18(a), (e), and (k) will respectively be named $H_0$,
$H_1$, and $H_2$.
It can be shown that the only primitive regge color factors entering into
SCC diagrams are $H_p\ (p=0,1,2,3,\:)$,  representing a
pattern with $p$ horizontal steps climbing up from left to right.

It can also be shown that regge color factors with
the same primitive components can be identified in the leading-log
approximation. For that reason a regge color factor can simply be labelled
by the number of each primitive component it contains, {\it i.e.,} can be
expressed in the form $\Phi=\prod_{p=0}^\infty H_p^{f_p}$. For example, the
color factors of 18(b), (c), (d) are $H_0^2, H_0^3$, and $H_0^4$ respectively.
The others are $18(f)=18(g)=H_0H_1, 18(h)=H_0^2H_1, 18(i)=18(j)=H_1^2$.

The next step is to collect all SC diagrams contributing to a given
regge color factor $\Phi$. It turns out that these diagrams can be
summed up using the factorization theorem (3.3) to be
($\A=-T/2s$)
\be
\A\{\Phi\}(\Delta)=\prod_{p=0}(-i)^{f_p-1}\[*\A\{H_p\}\]^{f_p}(\Delta)\ ,\ee
where $\A\{H_p\}$ is the sum of all SC amplitudes whose corresponding SCC
diagrams contain the connected primitive color factor $H_p$. This shows
that the reggeized factorization hypothesis holds for all $s$-channel-ladder
diagrams.

In particular,
$\A\{H_0\}(\Delta)$ is simply $g^2R_1(s,\Delta)$, where $R_1$ is given
by (5.5) to order $g^2$. The $g^4$ contribution comes from diagrams $C1_c$
and $C2_c$ of Fig.~17 and they are not SC diagrams.

Using (5.6), the ${\bf G}_2$ and ${\bf G}_4$ amplitudes of (5.7) are special
cases of (6.3) with $f_0=2,3$ and all other $f_p=0$.

Similarly, $\A\{H_1\}(\Delta)=(-i/2)R_{2,1}(s,\Delta)$, where $R_{2,1}$ is
given by the first term of (5.8) -- the second term there does not come from
SC diagrams.

\bigskip\bigskip
\n{\bf Acknowledgements}

\medskip
This research is supported in part by the by the Natural Science and
Engineering Research Council of Canada, and the Fonds pour la
Formation de Chercheurs et l'Aide \`a la Recherche of Qu\'ebec.
I would like to thank my collaborators Yong-Jian Feng, Omid Hamidi-Ravari,
and Keh-Fei Liu. I am specially grateful to Yong-Jian Feng for drawing
the extra diagrams in these notes.


\begin{thebibliography}{99}
\bibitem{1} C.S. Lam and K.F. Liu, McGill/96-12=hep-ph/9604377, to appear
in {\it Nucl. Phys. B}.
\bibitem{2} Y.J. Feng, O. Hamidi-Ravari, and C.S. Lam,
{\it Phys. Rev. D} {\bf 54} (1996) 3114 (McGill/96-13=hep-ph/9604429).
\bibitem{3} G. 't Hooft, {\it Nucl. Phys.} {\bf B72} (1974) 461;
E. Witten, {\it Nucl. Phys.} {\bf B160} (1979) 57;
S. Coleman, Erice Lectures (1979).
\bibitem{4}
R.F. Dashen, E. Jenkins and A.V. Manohar, {\it Phys. Rev. D}
{\bf 49} 4713;
M.A. Luty and J. March-Russell, {\it Nucl. Phys.}
{\bf B426} (1994) 71.
\bibitem{5} C.S. Lam, in `Proceedings of the International Symposium
on Heavy Flavor and Electroweak Theory', Aug.~1995, Beijing, 
{\it Eds: C.H. Chang and
C.S. Huang}, World Scientific (1996) 
(McGill/96-22=hep-ph/9606350); C.S. Lam and K.F. Liu,
to be published.
\bibitem{6} H. Cheng and T.T. Wu, `{\it Expanding Protons: Scattering
at High Energies}', (M.I.T. press, 1987).
\bibitem{7} H. Cheng and T.T. Wu, Phys.~Rev. 186 (1969) 1611;
M. Levy and J. Sucher, Phys.~Rev. 186 (1969) 1656.
\bibitem{8} C.Y. Lo and H. Cheng,
Phys.~Rev. D13 (1976) 1131; D15 (1077) 2959; H. Cheng, J.A. Dickinson, and
K. Olaussen, Phys.~Rev. D23 (1981) 534.
\bibitem{9} L.N. Lipatov,  Yad.~Fiz. 23 (1976) 642 [Sov.~J.~Nucl.~Phys.
23 (1976) 338]; Ya. Ya. Balitskii and L.N. Lipatov, Yad.~Fiz. 28 (1978) 1597
[Sov.~J.~Nucl.~Phys. 28 (1978) 822];
E.A. Kuraev, L.N. Lipatov, and V.S. Fadin, Zh.~Eksp.~Teor.~Fiz.
71 (1976) 840 [Sov.~Phys. JETP 44 (1976) 443]; {\it ibid.} 72 (1977) 377 [{\it
ibid.} 45 (1977) 199]; L.N. Lipatov, in `Perturbative Quantum Chromodynamics'
(A.H. Mueller, ed., World Scientific 1989);
V. Del Duca, hep-ph/9503226.
\bibitem{10} Y.J. Feng and C.S. Lam, McGill/96-19=hep-ph/9606351.
\bibitem{11} C.S. Lam, McGill/96-23=hep-ph/9606374.
\bibitem{12} Y.J. Feng and C.S. Lam, Phys.~Rev. D50 (1994) 7430.
\bibitem{13} F.E. Low, Phys.~Rev. D12 (1975) 163;
S. Nussinov, Phys.~Rev.~Lett. 34 (1975) 1286.

\end{thebibliography}
\end{document}